\documentclass[aps,prd,twocolumn,showpacs]{revtex4} 
\usepackage{graphicx}

\font\FermiSmallfont=cmssq8 scaled 1200

\def\LANLppthead#1{
\null 
\begin{center}\vskip -1.0truein{\hbox to 7.5truein {
\hfill
\vbox to 1in {\vfill \FermiSmallfont
              \hbox{#1}
              \vfill}
}}\vskip-0.0truein\end{center}}

\begin{document}

\title{New Nuclear Physics for Big Bang Nucleosynthesis}
\author{Richard N.\ Boyd$^1$, Carl R.\ Brune$^2$, George M.\ Fuller$^3$, and Christel J.\ Smith$^4$}
\affiliation{$^1$Lawrence Livermore National Laboratory, Livermore, CA 94551\\
$^2$Department of Physics and Astronomy, Ohio University, Athens, OH 45701\\
$^3$Department of Physics, University of California, San Diego, La Jolla, CA 92093-0319\\
$^4$Physics Department, Arizona State University, Tempe, AZ 85287-1404}

\date{\today}

\begin{abstract}
We discuss nuclear reactions which could play a role in Big Bang Nucleosynthesis (BBN).  Most of these reactions involve lithium and beryllium isotopes and the rates for some of these have not previously been included in BBN calculations.  Few of these reactions are well studied in the laboratory.  We also discuss novel effects in these reactions, including thermal population of nuclear target states, resonant enhancement, and non-thermal neutron reaction products.  We perform sensitivity studies which show that even given considerable nuclear physics uncertainties, most of these nuclear reactions have minimal leverage on the standard BBN abundance yields of $^6$Li and $^7$Li.  Although a few have the potential to alter the yields significantly, we argue that this is unlikely.

\end{abstract}
\maketitle

\section{Introduction}

The basic details of big bang nucleosynthesis (BBN) have been well understood for several decades \cite{wag69, wfh, JYang, Walker:1991ap, 1990ApJ...358...47K, skm, kawano, kawano1, Nollett:2000fh,Schramm:1997vs,ourcode,coulfac,Olive,Steigman1,steig}, and nucleosynthesis predictions are in reasonable agreement with the abundances of $^2$H, $^3$He, and $^4$He observed in metal poor environments.  However, there are new challenges in achieving agreement between theory and observation for the lithium isotopes, $^6$Li and $^7$Li.  These challenges stem from recent observations of metal-poor halo stars \cite{asplund} and the high precision measurement of the baryon-to-photon ratio of the Universe by WMAP \cite{WMAP, WMAP1,3yrwmap, Bennett:2003ca, Komatsu:2010fb}.  The WMAP value, which confirms earlier inference of this quantity from deuterium \cite{Tytler, Omeara}, is likely to be improved on by future cosmic microwave background observations ({\it e.g.,} Planck \cite{planck}).

The abundance of $^7$Li has been a longstanding problem, as the predicted abundance is higher than observed in metal poor halo stars \cite{spite} by roughly a factor of three at the WMAP baryon-to-photon ratio. Attempts have been made to reconcile this difference, primarily on the basis of stellar processing of $^7$Li. Although it can be argued that some $^7$Li would be destroyed by the stars in which it is observed, it has been found to be very difficult to justify enough destruction to bring theory and observation into agreement \cite{1999ApJ...523..654R}. 

Recently Very Large Telescope observations \cite{asplund} suggested an abundance of $^6$Li on the surface of some metal-poor halo stars roughly $\sim 1/30$ of that of $^7$Li.  Although this observation has been challenged by other work \cite{2007A&A...473L..37C}, if it is correct, it represents a 2-3 order of magnitude discrepancy between theory and observation.  Much is riding on this discrepancy.  If $^6$Li must be made by post-BBN non-thermal processes, then the abundance of this nucleus conceivably may be an indirect probe of new physics, specifically heavy particle decay and dark matter \cite{1475-7516-2006-11-014,Jedamzik:1999di,1988PhRvL..60Q...7D}.

Nollett {\it et al.} \cite{Nollett:1996ef} have provided an insightful and comprehensive study of the nuclear reactions which bear on $^6$Li production and destruction, while Cyburt and Pospelov \cite{Cyburt:2009cf} have explored the necessary issues for a pure nuclear physics solution to the $^7$Li overproduction problem.  In this paper we explore related issues, find no clear paths to solution of either lithium problem, but do provide rates for incorporation into BBN codes as well as studies of the sensitivity of lithium isotope production and destruction to uncertainties in these rates.  It should be kept in mind that although these rates may make little difference in standard BBN abundance yields, using BBN to constrain non-standard scenarios with new particle physics may require new levels of precision in some parts of the BBN nuclear reaction network \cite{sfka,abfw}.

Although the standard hot BBN model code contains most of the reactions that could be relevant to BBN, those involving reactions on short-lived nuclei are not well studied, and in some cases have not been included. Given the current level of precision of BBN calculations \cite{Schramm:1997vs,Burles:2000zk,Steigman:2007xt}, it seems appropriate to reexamine the BBN network to be sure that all possible reactions are included, and to study the potential effects of those reactions for which data do not exist. Candidate reactions here include $^7$Be($^3$H,$^4$He)$^6$Li and, possibly, other reactions on $^7$Be, particularly because mass-7 is made primarily as $^7$Be at the WMAP baryonic density. Another reaction that might be of particular interest for synthesizing $^6$Li could be $^3$H($^3$He,$\gamma$)$^6$Li, which was measured, but in a very difficult experiment. 

Another aspect of BBN that we have studied is the inclusion of nonthermal particles. It has generally been assumed that the BBN reactions occur in an environment where ions have thermal, Maxwellian velocity distributions.  This approximation could be violated at some level by reactions induced by highly-energetic particles produced by exothermic reactions. Again, in the interest of studying the precision that can be expected of BBN calculations\cite{Schramm:1997vs,Burles:2000zk,Steigman:2007xt}, we have studied the effects of non-thermal particles. The most likely scenarios for this process involve neutrons produced in $^3$H($d,n$)$^4$He and the protons produced in $^3$He($d,p$)$^4$He. These reactions involve relatively abundant nuclei in the entrance channel, have large cross sections, and produce nuclei with energies in excess of 10 MeV. Some aspects of this were studied recently \cite{Voronchev:2008zz}. That work showed that thermalization of the non-thermal charged particles occurs sufficiently rapidly, through electromagnetic processes, that those particles have only a tiny effect on the calculated BBN abundances. However, that study only investigated the thermalization of protons and did not consider neutron thermalization, which occurs over a much longer timescale than for protons. It might be thought that the non-thermal neutrons could affect the BBN abundances, specifically that of $^7$Be, through $^3$He($n,p$)$^3$H and through neutron-induced reactions on $^7$Be. The former reaction could reduce the abundance of $^3$He, and therefore, of $^7$Be, since it is made from $^3$He($^4$He,$\gamma$)$^7$Be. The latter reactions might reduce the abundance of $^7$Be after it was made, and $^7$Be($n,^2$H)$^6$Li, which is endothermic, could produce $^6$Li. Therefore we have considered the effects of non-thermal neutrons on the $^7$Be abundance, and on possible breakup of $^2$H as well.

The lowest-lying first excited states of the lightest nuclei are those of the mirror nuclei $^7$Li and $^7$Be.  The excitation energies of these excited states are within a factor of 3 or 4 of relevant BBN thermal energies.  Accordingly, we have examined how potentially key nuclear reactions like $^7$Be($d,\gamma)^9$B, $^7$Be($d,p)2\alpha$, and $^7$Be($d,^3$He)$^6$Li could be altered if they were to proceed through a thermally populated first excited state, $^7$Be*.  Cyburt, Ref.~\cite{Cyburt:2009cf}, argued that resonant enhancement of $^7$Be($d,p)2\alpha$ could in principle solve the first of the lithium problems outlined above, potentially reducing the $^7$Li BBN yield by a factor of 3 or so.  We revisited that possibility, but also find that if this reaction were to proceed through $^7$Be* a different resonant enhancement channel can come into play.  In the end, however, standard BBN yield alterations are found to be small due to the small thermal population of $^7$Be*.

In what follows we discuss the BBN reaction network in section II and the results of our BBN calculations with the new rates in section III.  In section IV we discuss non-thermal neutron effects and in section V we discuss thermal population of excited states and resonant enhancement.  We give conclusions in section VI.

\section{BBN Nuclear Reaction Network}
\begin{figure*}
\includegraphics[trim=0in 0in 1in 1.2in,clip=true,width=5.75in,angle=90]{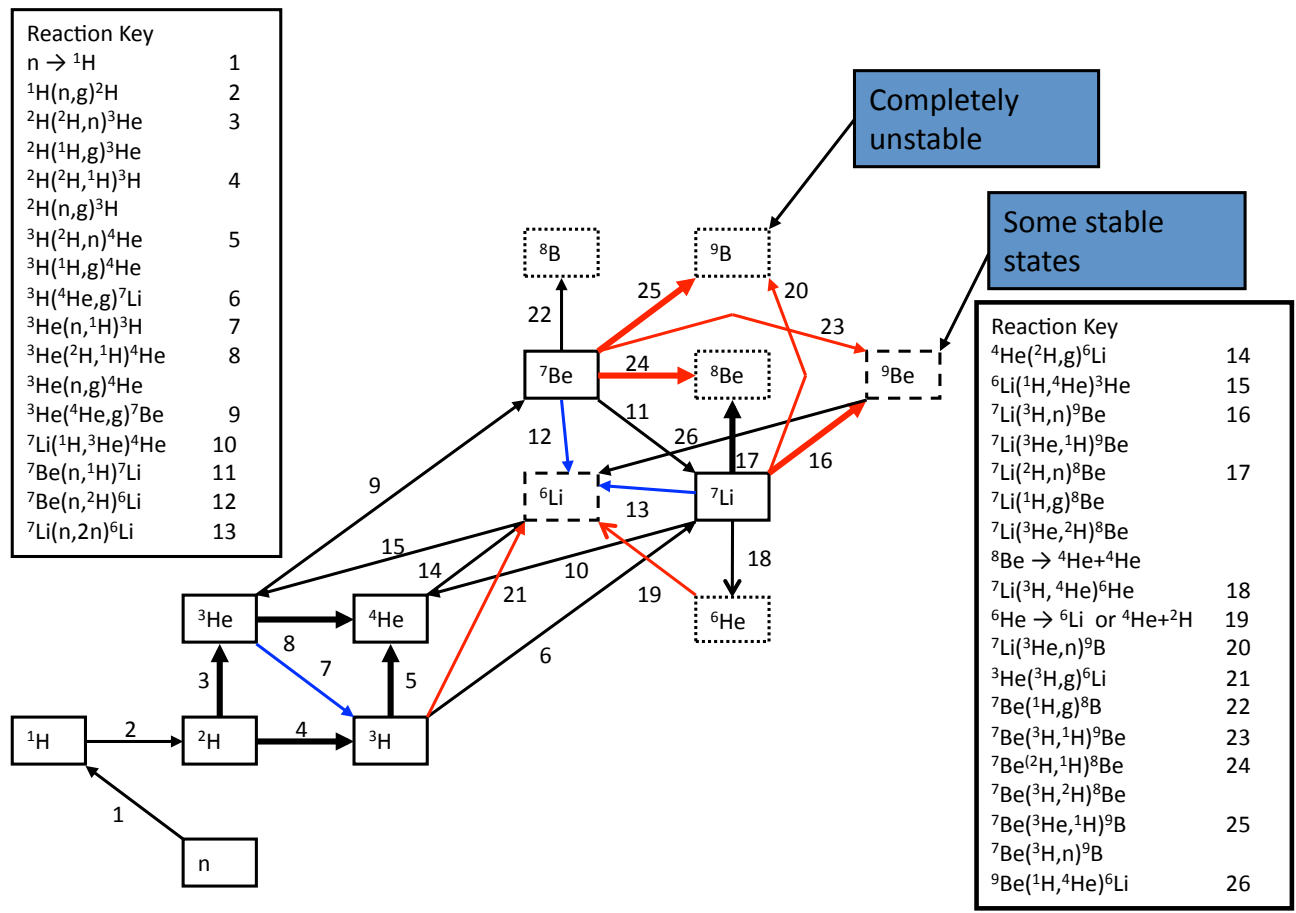}
\caption{Reaction network for BBN, as modified from \cite{boydbook}. The nuclei indicated in dotted boxes are completely unstable, whereas those in dashed boxes have some stable states and some others that are relevant to BBN that undergo particle decay to some other nuclei. Red lines indicate reactions newly added to the BBN code, and blue lines indicate reactions studied in the context of non-thermal neutrons. Double thickness lines indicate more than one possible reaction.}
\label{network}
\end{figure*}

In this paper, we discuss several reactions that should naturally be a part of the BBN reaction network, but have not generally been included. In most cases this is because these reactions are difficult to study, for example, because either the target nucleus or the beam nucleus, or in some cases both, are radioactive. The inputs to the BBN code, that is, the information from the experiments, is in the form of thermonuclear reaction rates, $\langle\sigma v\rangle$, defined as \cite{boydbook}
\begin{eqnarray}
\langle\sigma v\rangle = \left[{8\over{\pi\mu}}\right]^{1/2}[kT]^{-3/2}\int\sigma(E) E e^{-E/kT} dE,\\
\nonumber
\label{sigmav}
\end{eqnarray}
where $\mu$ is the reduced mass and $E$ is the center of mass energy of the reactants. Nuclear astrophysical measurements usually produce an astrophysical S-factor, $S(E)$, defined as
\begin{eqnarray}
S(E)=E\sigma(E) e^{bE^{-1/2}},\\
\nonumber
\label{astrofac}
\end{eqnarray}
where $e^{-bE^{-1/2}}$ is (approximately) the Coulomb barrier tunneling probability, $b = (2\mu)^{1/2} \pi e^2 z_1Z_1/\hbar$, and $z_1$ and $Z_1$ are the charge numbers of the entrance channel nuclei. In the absence of resonances, $S(E)$ will be fairly constant, and the thermonuclear reaction rate $\langle\sigma v\rangle$ times AvagadroÕs constant, $N_A$, can be shown \cite{boydbook} to be: 
\begin{equation}
N_A\langle\sigma v\rangle = 4.34\times10^5 {{S(E)}\over{ [\mu z_1 Z_1]}} \tau^2 e^{-\tau} {{{\rm cm}^3}\over{ {\rm s}-{\rm mole}}},	
\label{Nsig}
\end{equation}
where $\tau = 4.246 [z_1^2 Z_1^2 \mu / T_9]^{1/3}$, with $T_9$ being the temperature in billions Kelvin, and $S(E)$ is in units of keV-barns. Since most of the changes we made to the BBN reaction rates involved reactions that produced high excitation energies in the compound nucleus, we found that the S-factors we calculated were usually sufficiently non-resonant that we could use the above prescription. 

\begin{center}
\begin{table*}
\caption{}
\begin{tabular}{ | l | r | l | l |}
\multicolumn{4}{c}{Reactions Added to the BBN Code} \\
\hline
Reaction&Q-value (MeV) & Comment & Effect\\
\hline\hline
$^7$Li($^3$H,$n)^9$Be(ground state)&10.439&1 (Ref. \cite{Brune:1991zza})&none\\
$^7$Li($^3$H,$n)^9$Be(excited states)&10.439&1 (Ref. \cite{Brune:1991zza})&	none\\
$^7$Li($^3$He,$p)^9$Be(ground state)&11.202&2 (Ref. \cite{Rath1990338,jyanphd})&none\\
$^7$Li($^3$He,$p)^9$Be(excited states)&11.202&2 (Ref. \cite{Rath1990338,jyanphd})&none\\
$^7$Li($^3$He,$n)^9$B&9.352&9 (Ref. \cite{Duggan1963336})&none\\
$^7$Li($^3$He,$^2$H)$^8$Be&17.608&8 (Ref. \cite{pallone, Warner1987339})&none\\
$^7$Li($^2$H,$n)^8$Be&15.031&8 (Ref. \cite{pallone, Warner1987339})&none\\
$^7$Li($^3$He,$^4$He)$^6$Li(ground state)&13.328&4 (Ref. \cite{Forsyth1965517})&none\\
$^7$Li($^3$He,$^4$He)$^6$Li(excited states)&13.328&4 (Ref. \cite{Forsyth1965517})&none\\
$^7$Li($^3$H,$^4$He)$^6$He&9.838&3 (Ref. \cite{PhysRevC.27.6})&none\\
$^7$Be($^3$H,$^4$He)$^6$Li(ground state)&14.208&	4&small\\
$^7$Be($^3$H,$^4$He)$^6$Li(excited states)&14.208&	4&small\\
$^7$Be($^3$H,$p)^9$Be(ground state)&12.082&6&small\\
$^7$Be($^3$H,$p)^9$Be(excited states)&12.082&6&small\\
$^3$He($^3$H,$\gamma)^6$Li(low lying states)&15.795&5 (Ref. \cite{PhysRevLett.25.1764, Blatt:1968zz})&large\\
$^9$Be($p,\alpha)^6$Li(ground state)&2.126&Ref. \cite{NACRE}&none\\
$^7$Be($p,\gamma)^8$B&0.137&Ref. \cite{NACRE}&none\\
$^7$Be($^2$H,$p)^8$Be&16.674&8 (Ref. \cite{pallone, Warner1987339})&large\\
$^7$Be($^3$H,$^4$He)$^6$Li&14.208&	4&small\\
$^7$Be($^3$H,$n)^9$B&10.232&7&small\\
$^7$Be($^3$H,$^2$H)$^8$Be	&12.641&8 (Ref. \cite{pallone, Warner1987339})&small\\
$^7$Be($^3$He,$p)^9$B&10.995&7&none\\
$^7$Be($^2$H,$^3$He)$^6$Li&-0.112&10&none\\
$^7$Be*($^2$H,$^3$He)$^6$Li&0.317&10&large\\
$^7$Be*($^2$H,$p)^8$Be&17.103&8 (Ref. \cite{pallone, Warner1987339})&large\\
$^7$Be*($p,\gamma)^8$B&0.566&Ref. \cite{NACRE}&large\\

\hline
\end{tabular}
\end{table*}
\end{center}

In one case we needed to calculate the reaction rate for a narrow resonance, for which the formula is \cite{boydbook}
\begin{widetext}
\begin{equation}
N_A\langle\sigma v\rangle = 1.535\times10^{12} (\omega\gamma)_3 \mu^{-3/2} T_9^{-3/2} e^{-11.605 E_6 / T_9}\ {\rm cm}^3\ {\rm s}^{-1}\  {\rm mole}^{-1},   
\label{narrowres}
\end{equation}
\end{widetext}
where $(\omega\gamma)_3=\omega\Gamma_1\Gamma_2/\Gamma$ (given in keV), with $\Gamma_1$ and $\Gamma_2$ the resonance partial widths in the entrance and exit channels, $\Gamma$ is the total width, and $\omega$ is the spin statistics factor = $(2J+1)/[(2J_1+1)(2J_2+1)]$, with $J$ the total resonance spin and $J_1$ and $J_2$ the spins of the two entrance channel particles.  $E_6$ is the center of mass resonance energy in the entrance channel in MeV. This formula does not apply to cases in which there are broad resonance structures; then the reaction rates need to be calculated by numerical integration or approximated in other ways.

The study by Serpico $et$ $al.$ \cite{Serpico:2004gx} investigated the sensitivity of BBN abundance yields to most of the usually included reactions. We have added several \lq\lq deuteron transfer\rq\rq\ reactions.  These can be quite strong, and conceivably could have an appreciable effect on the $^7$Li and $^7$Be abundances. For example, the reaction rate for $^7$Li($^3$H,$n$)$^9$Be, measured by Brune {\it et al.} \cite{Brune:1991zza}, was found to proceed to both the $^9$Be ground state and to excited states.  Since the rate to the ground state is 0.084 times the rate to the continuum states, destruction of $^7$Li is far more likely than production of $^9$Be by this reaction. We have also added to the code several reactions that occur on $^7$Be, {\it e.g.}, $^7$Be($p,\gamma)^8$B \cite{NACRE}, $^7$Be($d,p)^8$Be and $^7$Be($^3$H,$n)^9$B \cite{Duggan1963336}, all of which create nuclides that undergo particle decay, and so result in destruction of $^7$Be. 

Another reaction that has interesting potential is $^3$He($^3$H,$\gamma)^6$Li, which has been studied experimentally \cite{Blatt:1968zz,PhysRevLett.25.1764}, has a huge positive Q-value (15.795 MeV), and can go to many excited states of $^6$Li. The $^6$Li ground state is stable, the second excited state decays to the ground state, and all other states undergo breakup to $^4$He+$^2$H. Thus this reaction can produce some $^6$Li. Its effects were studied theoretically by Fukugita and Kajino \cite{Fukugita:1990kj}, and by Madsen \cite{PhysRevD.41.2472}, and were found to contribute little to $^6$Li production. However, there is some question as to the normalization of the cross section for that reaction; this will be discussed below. Another promising reaction is $^7$Li($^3$He,$^4$He)$^6$Li \cite{Forsyth1965517}, which proceeds through high-lying resonances in $^{10}$B, and may be able to produce some $^6$Li. Finally, $^7$Be($^3$H,$^4$He)$^6$Li can both destroy mass-7 nuclei and produce $^6$Li. This reaction has not been studied experimentally, since both nuclei in the entrance channel are radioactive. It has a large Q-value, so high-lying states in $^6$Li that it can populate must be considered in the context of mass-7 destruction.

We have terminated our reaction network at mass 9. We believe this is justified by the $^9$Be($p,\alpha)^6$Li reaction rate \cite{NACRE}; its magnitude insures that nearly all of the $^9$Be made by $^7$Li($^3$H,$n$)$^9$Be  or any other reaction will be returned to $^6$Li. As pointed out by Boyd and Kajino \cite{1989ApJ...336L..55B}, $^9$Be($^3$H,$n)^{11}$B also has the potential to destroy $^9$Be, but its rate is so much smaller than that for $^9$Be($p,\alpha)^6$Li \cite{NACRE} that it is of little consequence for destroying $^9$Be, and results in so little production of mass-11 nuclei that they will be far below observational limits.

In the end, however, our attempts to augment the production of $^6$Li were undone, principally by $^6$Li($p,\alpha)^3$He, the large cross section for which results in the destruction of nearly all of the $^6$Li made in BBN. ÊThis was nicely set out in Nollett $et$ $al.$ \cite{Nollett:2000fh}.  In the next section we will add to this discussion.

\subsection{Comments on the Added Nuclear Reaction Rates}

Some comments are in order concerning the reaction rates we used in our calculations.  These reactions are listed in Table I.

\subsubsection{$^7${\rm Li($^3$H,$n)^9$Be}(ground state) and {\rm $^7$Li($^3$H,$n)^9$Be}(excited states)} 

These rates have been determined experimentally \cite{Brune:1991zza}, so are known accurately. The reaction to the ground state gives one neutron, whereas those to all the excited states give two neutrons due to the breakup of all $^9$Be excited states into two $\alpha$-particles and a neutron. The relevant rates Ê\cite{Brune:1991zza} are:
\begin{eqnarray}
&N_A\langle\sigma v(^7{\rm Li}(^3{\rm H},n)^9{\rm Be(ground\ state)})\rangle =\\
&2.98\times10^{10} T_9^{-2/3} e^{-11.327 T_9^{-1/3}}\nonumber\\
&\times [1- 0.122T_9^{2/3} + 1.32(T_9^{4/3} Ð 0.127T_9^{2/3} + 0.0742)^{-1}]\nonumber\\
&{\rm cm}^3\ {\rm s}^{-1}\ {\rm mole}^{-1},\nonumber\\
\nonumber\\
&N_A\langle\sigma v(^7{\rm Li}(^3{\rm H},n)^9{\rm Be(excited\ states)})\rangle = \\
&11.9\times N_A\langle\sigma v(^7{\rm Li}(^3{\rm H},n)^9{\rm Be(ground\ state)})\rangle.\nonumber
\end{eqnarray}

\subsubsection{{\rm $^7$Li($^3$He,$p)^9$Be} (ground state) and {\rm $^7$Li($^3$He,$p)^9$Be} (excited states)}

The first reaction was studied at higher energies \cite{Rath1990338}, and extended to lower energies \cite{jyanphd}, and an S-factor was determined for the reaction to the $^9$Be(ground state).  Although the S-factor exhibits a broad resonance structure, the resonances exist at relatively high energies.  At energies below 100 keV, where this reaction is relevant to BBN, the S-factor is essentially flat at 5.3 MeV-barn.  In addition, from the spectrum \cite{Rath1990338} it can be inferred that the reaction to excited and continuum states is roughly a factor of five greater than that to ground state.  Therefore the assumed rates, in units of cm$^3$ s$^{-1}$ mole$^{-1}$, are:

\begin{eqnarray}
&N_A\langle\sigma v(^7{\rm Li}(^3{\rm He},p)^9{\rm Be (ground\ state}))\rangle = \\
&5.87\times10^{10}T_9^{-2/3}e^{-17.980 T_9^{-1/3}}\nonumber\\
\nonumber\\
&N_A\langle\sigma v(^7{\rm Li}(^3{\rm He},p)^9{\rm Be (excited\ states}))\rangle  = \\
&2.94\times10^{11}T_9^{-2/3}e^{-17.980 T_9^{-1/3}}\nonumber
\end{eqnarray}

\subsubsection{{\rm $^7$Li($^3$H,$^4$He)$^6$He}}

Although this reaction makes $^6$He, its ground state $\beta$-decays with a half-life of 807 ms to $^6$Li. It was studied at low energies \cite{PhysRevC.27.6}, both to the ground state, and to the first excited state, which breaks up into $^4$He and two neutrons.  The S-factor to the $^6$He ground state was estimated from data in Ref.~\cite{PhysRevC.27.6} to be 0.84 MeV-barn, and that to the first excited state to be 13.9 MeV-barn. For the two rates associated with this reaction, this gives
\begin{eqnarray}
&N_A\langle\sigma v(^7{\rm Li}(^3{\rm H},^4{\rm He})^6{\rm He(ground\ state}))\rangle =\\
&7.4\times10^9 T_9^{-2/3} e^{-11.327 T_9^{-1/3}}\ {\rm cm}^3\ {\rm s}^{-1}\ {\rm mole}^{-1},\nonumber\\
\nonumber\\
&N_A\langle\sigma v(^7{\rm Li}(^3{\rm H},^4{\rm He}+2n)^4{\rm He})\rangle = \\
&1.22\times10^{11} T_9^{-2/3} e^{-11.327 T_9^{-1/3}}\ {\rm cm}^3\ {\rm s}^{-1}\ {\rm mole}^{-1}.\nonumber
\end{eqnarray}

\subsubsection{{\rm $^7$Li($^3$He,$^4$He)$^6$Li} and {\rm $^7$Be($^3$H,$^4$He)$^6$Li}}

The first reaction was studied at fairly low energies \cite{Forsyth1965517}.  Data suggest a fairly constant S-factor for the ground state over almost 1 MeV in the center of mass. Data for the ground state and the first excited state (which decays to $^2$H$+^4$He) are well established, and an approximate total cross section for the second excited state ($T=1$, so is narrow, and decays to the $^6$Li ground state) can be inferred. Adding the S-factors for $^7$Li($^3$He,$^4$He)$^6$Li to the ground state (17.2 MeV-barn) and second excited state (12.3 MeV-barn, taken at the lowest energy) gives 29.5 MeV-barn. In the absence of data for the second reaction, we have assumed that the same S-factors for making $^6$Li apply to it, as well as for destroying $^7$Li or $^7$Be. The first excited state is excited with about 5/7 of the strength of the ground state in the $^7$Li($^3$He,$^4$He) reaction. Therefore the assumed rates, all in units of cm$^3$ s$^{-1}$ mole$^{-1}$, are:
\begin{eqnarray}
&N_A\langle\sigma v(^7{\rm Li}(^3{\rm He},^4{\rm He})^6{\rm Li(ground\ state}))\rangle=\\
&3.27\times10^{11} T_9^{-2/3} e^{-17.980 T_9^{-1/3}},\nonumber\\ 
\nonumber\\
&N_A\langle\sigma v(^7{\rm Li}(^3{\rm He},^4{\rm He})^6{\rm Li(1^{st}\ excited\ states}))\rangle=\\
&1.36\times10^{11} T_9^{-2/3} e^{-17.980 T_9^{-1/3}},\nonumber\\
\nonumber\\
&N_A\langle\sigma v(^7{\rm Be}(^3{\rm H},^4{\rm He})^6{\rm Li(ground\ state}))\rangle=\\
&= 2.87\times10^{11} T_9^{-2/3} e^{-13.722 T_9^{-1/3}},\nonumber\\
\nonumber\\
&N_A\langle\sigma v(^7{\rm Be}(^3{\rm H},^4{\rm He})^6{\rm Li(1^{st}\ excited\ states}))\rangle=\\
&1.19\times10^{11} T_9^{-2/3} e^{-13.722 T_9^{-1/3}}.\nonumber
\end{eqnarray}

\subsubsection{\rm {$^3$He($^3$H,$\gamma)^6$Li}}
Despite the difficulty of this experiment (primarily because of a huge neutron background), this cross section both to the low lying states and to higher lying states has been measured \cite{Blatt:1968zz}. The $^6$Li ground state is populated, and the second excited state, which also decays to the ground state, has a cross section of about 18\% of that to the ground state. However, the S-factor is difficult to ascertain from the data for this reaction. It appears to increase linearly from near zero at low energy to a much larger value before it levels off \cite{Fukugita:1990kj,PhysRevD.41.2472}. Since this reaction is thought \cite{PhysRevLett.25.1764} to result from continuum capture, this is surprising. A more likely explanation is that the lower energy points distort the results (the beam was stopped in the target at the lower energies, complicating the determination of the effective target thickness). We therefore calculated the S-factor for the points at around 1 MeV in the center of mass, and assumed it to be constant, as might be expected if the reaction is indeed continuum capture. The value for the ground state was increased by 18\% to accommodate the contribution to $^6$Li formation from the second excited state. The resulting reaction rate, in units of cm$^3$ s$^{-1}$ mole$^{-1}$, is:
\begin{eqnarray}
&N_A\langle\sigma v(^3{\rm He}(^3{\rm H},\gamma)^6{\rm Li})\rangle=\\
&1.02\times10^7 T_9^{-2/3} e^{-7.729 T_9^{-1/3}}.\nonumber
\end{eqnarray}

\subsubsection{\rm {$^7$Be($^3$H,$p)^9$Be}}

The dominant feature of this reaction appears to be a broad resonance in $^{10}$B \cite{jyanphd}, $\Gamma$ Ê= 600 keV, at 18.800 MeV, seen in several reactions, which is 132 keV above the $^3$H$+^7$Be threshold. Since this is a broad resonance, much of which is sub-threshold, it would not be expected to contribute strongly to this reaction. Accordingly, in the absence of better information, we have simply assumed the same rate for this reaction as for the $^7$Be($^3$H,$^4$He)$^6$Li reaction (to the excited states). Because this resonance would be expected to break up into $^4$He+$^4$He+$^2$H most of the time (and its decay to $^6$Li is included in item 4), the rate for this reaction only destroys $^7$Be (in units of cm$^3$ s$^{-1}$ mole$^{-1}$):
\begin{eqnarray}
&N_A\langle\sigma v(^7{\rm Be}(^3{\rm H},\alpha+n)^4{\rm He})\rangle=\\
&1.19\times10^{11} T_9^{-2/3} e^{-13.722 T_9^{-1/3}}.\nonumber
\end{eqnarray}

\subsubsection{{\rm {$^7$Be($^3$He,$p)^9$B}} and {\rm $^7$Be($^3$H,$n)^9$B}}
These reactions go to a variety of states in $^9$B, all of which decay by proton emission, including the ground state. Since these are deuteron transfer reactions, and have comparable (large) Q-values to that for the $^7$Li($^3$H,$n)$ reaction, we assumed the same astrophysical S-factor for it to the $^9$B ground state as for $^7$Li($^3$H,$n)^9$Be(ground state) \cite{Brune:1991zza} (including more than the leading term does not seem justifiable), and noted \cite{Rath1990338} that the $^7$Li($^3$He,$p)^9$Be yield to all states appeared to be greater than that to the ground state by approximately a factor of five. After appropriate corrections for the nuclear charges \cite{boydbook}, the assumed rates are
\begin{eqnarray}
&N_A\langle\sigma v(^7{\rm Be}(^3{\rm He},p)^9{\rm B})\rangle=\\
&2.06\times10^{11} T_9^{-2/3} e^{-21.782 T_9^{-1/3}},\nonumber\\
\nonumber\\
&N_A\langle\sigma v(^7{\rm Be}(^3{\rm H},n)^9{\rm B})\rangle=\\
&1.64\times10^{11} T_9^{-2/3} e^{-13.722 T_9^{-1/3}},\nonumber
\end{eqnarray}
where $N_A\langle\sigma v\rangle$ is in the usual units of cm$^3$ s$^{-1}$ mole$^{-1}$.  Another reaction, $^9$Be($^3$He,$^6$Li)$^6$Li, was not included in the reaction network, as it was found  \cite{1969NuPhA.126..615T} to have a much smaller cross section than $^9$Be($p,^4$He)$^6$Li, and it depends on the $^9$Be abundance, which is small in BBN. Thus it would not be expected to contribute appreciably to $^6$Li abundance, despite the fact that it makes two $^6$Li nuclei, or to $^9$Be destruction. 

\subsubsection{{\rm {$^7$Li($^3$He,$d$)$^8$Be, $^7$Li($d,n)^8$Be, $^7$Be($d,p)^8$Be,}} and {\rm $^7$Be($^3$H,$d)^8$Be}}

These reactions are all single nucleon transfers, and all have large, and comparable, Q-values. We therefore estimated an S-factor for the $^7$Li($d,n)^8$Be reaction from the total cross section for that reaction \cite{pallone}, and applied it to all four reactions. That S-factor is reasonably well represented by a constant of about 2.5$\times10^4$ keV-barns (and which is lower than, but reasonably consistent with, that from Angulo {\it et al.} \cite{2005ApJ...630L.105A}, measured for $^7$Be($d,p)2^4$He); we therefore assumed that value, and then calculated the reaction rate from the standard expression \cite{boydbook}, including also spin statistics. Note that the $^7$Li($d,n)^8$Be reaction proceeds almost entirely through two states at 16.63 and 16.92 MeV in $^8$Be for energies in the compound system more than 1 MeV above the energy of those states \cite{2005ApJ...630L.105A}. However, the Q-values for all except $^7$Be($d,p)^8$Be are well below that energy, and even its Q-value is just at that suggested threshold \cite{national}. Thus the assumed rates, in units of cm$^3$ s$^{-1}$ mole$^{-1}$, are:
\begin{eqnarray}
&N_A\langle\sigma v(^7{\rm Li}(^3{\rm He},e)^8{\rm Be})\rangle=\\
&6.26\times10^{11} T_9^{-2/3} e^{-17.980 T_9^{-1/3}}\nonumber,\\
\nonumber\\
&N_A\langle\sigma v(^7{\rm Li}(d,n)^8{\rm Be})\rangle=\\
&2.43\times10^{11} T_9^{-2/3} e^{-10.254 T_9^{-1/3}}\nonumber,\\
\nonumber\\
&N_A\langle\sigma v(^7{\rm Be}(d,p)^8{\rm Be})\rangle=\\
&2.67\times10^{11} T_9^{-2/3} e^{-12.422 T_9^{-1/3}}\nonumber,\\
\nonumber\\
&N_A\langle\sigma v(^7{\rm Be}(^3{\rm H},d)^8{\rm Be})\rangle=\\
&5.47\times10^{11} T_9^{-2/3} e^{-13.722 T_9^{-1/3}}\nonumber.
\end{eqnarray}

A recent paper by Cyburt and Pospelov \cite{Cyburt:2009cf} suggests that there might be a resonance in the $^7$Be$+^2$H reaction that would decay to a variety of final states, all of which would destroy $^7$Be, and that this might solve the problem of strong $^7$Be production in BBN. This assertion was based on a paper by Dixit et al. [36], which studied $^9$Be$(p,pÕ)$, and which identified a strongly excited, and relatively narrow, state at 16.7 MeV. In that paper the possible structure of a narrow state at such high excitation is discussed, and is identified with earlier theoretical work [37] in which this state is suggested as having a structure of an $s_{1/2}$ neutron coupled to highly excited $2^+$ core states in $^8$Be. Such states could only be $\nu_{sd}^2 \nu_p^{-2}$, $\pi_{sd}^2 \pi_p^{-2}$, or $\nu_{sd}\pi_{sd}\nu_p^{-1}\pi_p^{-1}$ configurations, so would require a complicated (and therefore weak) multi-step reaction to excite any reactions involving $^2$H+$^7$Be, since $^7$Be presumably has a dominant configuration of $\nu_p^{-1}$ on an $^8$Be core.  As a consequence, it seems unlikely that reactions proceeding through this state could solve the $^7$Be excess problem.  Nonetheless, the considerable leverage that this reaction can have on $^7$Be, as pointed out by Cyburt and Pospelov \cite{Cyburt:2009cf}, remains alluring.

\subsubsection{\rm {$^7$Li($^3$He,$n)^9$B}}
Data for this reaction are sparse \cite{Duggan1963336}, but suggest that many states in $^9$B are populated. A spectrum at 2 MeV suggests that the strength to the ground state of $^9$B constitutes roughly 20\% of the yield (which, because the Q-value is so large, we took to be constant with 
energy). A rough S-factor can be obtained from the data for the transition to the ground state; we found it to be 180 MeV-barn, although this determination is complicated by an apparent resonance structure above the region of interest and a lack of information about experimental details. This gives a reaction rate, in units of cm$^3$ s$^{-1}$ mole$^{-1}$, Êof
\begin{eqnarray}
&N_A\langle\sigma v(^7{\rm Li}(^3{\rm He},e)^8{\rm Be})\rangle=\\
&2.0\times10^{12} T_9^{-2/3} e^{-17.980 T_9^{-1/3}}.\nonumber
\end{eqnarray}

\subsubsection{{\rm {$^7$Be($d,^3$He)$^6$Li}} and {\rm $^7$Be*($d,^3$He)$^6$Li}}

There are no data for the first reaction $^7$Be($d,^3$He)$^6$Li.   The Q-value for this reaction is slightly negative. However, if $^7$Be is in its first excited state at 0.429 MeV, which would be populated to some extent in the high temperature thermal environment in which BBN occurs, this situation changes. Then a strong resonance, with a width of 22 keV, would be expected at an excitation energy of 17.190 MeV in the $^9$B compound nuclear system, which is seen in the $^6$Li($^3$He,$p)$ reaction \cite{0954-3899-29-2-317}. In the absence of information about the partial widths of this state, we have assumed, in order to obtain an upper limit on its effect, that the partial widths for the $^6$Li$+^3$He and $^7$Be$+d$ channels were equal at 11 keV. Since the spin of the resonance is also not known, we have taken the spin statistical factor in Eq.~(\ref{narrowres}) to be 1.0. Then the reaction rate for this reaction is found to be:
\begin{eqnarray}
&N_A\langle\sigma v(^7{\rm Be*}(d,^3{\rm He})^6{\rm Li})\rangle=\\
&4.31\times10^{8} T_9^{-3/2} e^{-1.8336/T_9},\nonumber
\end{eqnarray}
 where $N_A\langle\sigma v\rangle$ is in the usual units of cm$^3$ s$^{-1}$ mole$^{-1}$.  In the BBN reaction network and abundance calculations the first excited state $^7$Be*
 was entered as a separate nuclear species.  The abundance of this species was tied to the abundance of the ground state by multiplying by the thermal population factor
\begin{equation}
P_{*}={{(2J^*+1)\cdot e^{-E_*/T}}\over{Z_{^7{\rm Be}}}}\approx {{2\cdot e^{-4.98/T_9}}\over{4}},
\end{equation}
where $J^*=1/2$ and $E^*\approx 0.4292$ MeV are the $^7$Be first excited state spin and excitation energy, respectively, and where we approximate the $^7$Be nuclear partition function as $Z_{^7{\rm Be}}\approx 2J_{\rm ground}+1=4.$  At $T_9\approx 1$, where $^7$Be is abundant enough to be affected appreciably by nuclear destruction reactions, this population factor is small, $P_*\approx 3\times10^{-3}$.  
 
\section{BBN Results from Added Nuclear Reactions}

We have employed a variant of the standard BBN reaction code \cite{skm}.  Our code is described in detail in Refs. \cite{ourcode, coulfac}.  The reaction network, which includes the standard model nuclei and reactions plus the reactions we have added, is indicated in Fig. \ref{network}. Note, however, that this cartoon does not show all the reactions in the network, but rather is simply a guide to some of the new reactions included here, including the inclusion of $^6$He to the network.  The reactions that were added to the BBN code, along with their Q-values, are listed in Table I. 

The evolution of the abundances in our calculations is shown in Fig. 2. The importance of the various reactions to the BBN abundances are specified below and indicated in Table 1. In general, it can be concluded that the added reactions have very little effect on BBN. However, since some of the reaction rates were quite uncertain, we increased the more uncertain rates by a factor of 1000 to see if such huge increases in the rates might have an effect. In most cases, only tiny effects were observed; the designation \lq\lq none\rq\rq\ means that the BBN abundances were not changed by more than 0.1\% with the factor of 1000 enhancement for that rate.  In this table, \lq\lq small\rq\rq\ means there was an alteration in abundance yield by more than $0.1\%$ when the rate was enhanced by a factor of 1000.  Likewise, in this table \lq\lq large\rq\rq\ means a 10\% or bigger abundance change when multiplying our best guess rate by 1000.

\begin{figure}
\includegraphics[width=3.1in,angle=270]{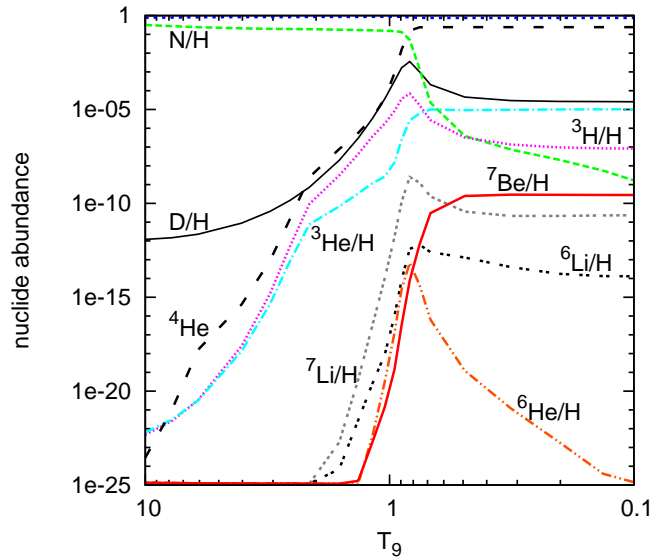}
\caption{Nuclear abundance as a function of temperature $T_9$, where $T_9=T/10^9$ K.  Abundances are given as mass fraction for $^4$He and number abundance relative to hydrogen for all others.}
\label{abun}
\end{figure}

It is remarkable that virtually nothing can be done in the context of the nuclear reactions to increase the $^6$Li abundance up to the value suggested by the observations, even with the huge multiplicative factors used, although the $^3$H($^3$He,$\gamma)^6$Li reaction might warrant further experimental study. Only the $^7$Be($^2$H,$^3$He)$^6$Li reaction produced a large enough effect to suggest that it might be important in destroying some of the mass-7 nuclei ultimately resulting from BBN. However, that required a factor of x1000 increase in the reaction rate, so that effect must be viewed with skepticism.

\begin{figure}
\includegraphics[width=3.1in,angle=270]{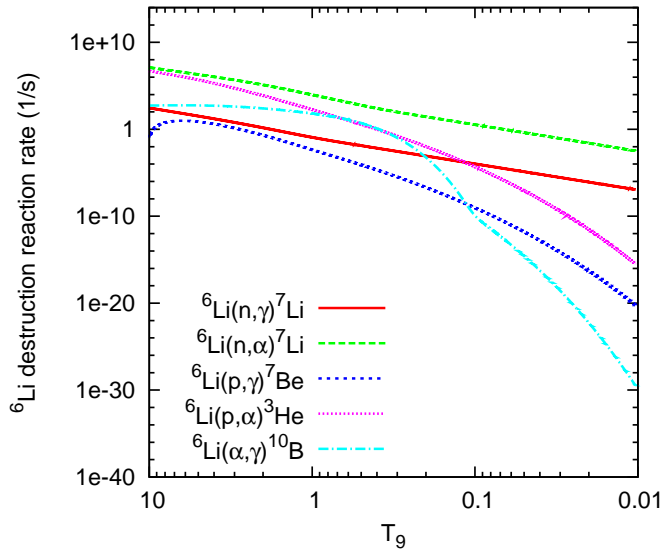}
\caption{$^6$Li destruction rates as a function of temperature $T_9$.}
\label{6Li}
\end{figure}

Of some interest was the inclusion of $^6$He in the reaction network. This species is converted quickly to $^6$Li by weak interactions.  Although the $^6$He total weak decay rate is accelerated by $\nu_e$ and $e^+$ capture processes at high temperature, by $T_9\sim 1$ the laboratory beta decay rate ($\sim 0.9\ {\rm s}^{-1}$) dominates \cite{lepcap}. Despite the rapid weak decay, as can be seen in Fig.~\ref{abun}, a significant $^6$He abundance is built up during BBN. ÊAlthough this $^6$He production initially contributes to $^6$Li through decay and lepton capture reactions \cite{lepcap}, the $^6$Li destruction processes effectively demolish any potential increase in the final $^6$Li abundance. ÊAll of the $^6$Li destruction reactions included in the BBN code are shown in Fig.~\ref{6Li}. ÊÊOf these processes, we see that $^6$Li($p,\alpha)^3$He has the largest effect in $^6$Li destruction mainly because of the abundance of protons when $^6$Li is created.

Following the Cyburt-Pospelov \cite{Cyburt:2009cf} suggestion on the sensitivity of the $^7$Be destruction to new nuclear physics, we performed simulations of BBN in which the rate for $^7$Be$(d,p)^8$Be was used as a surrogate for all the deuteron-induced reactions that destroy $^7$Be. ÊWe increased that rate by factors of 10 and 100 from the standard rate. The factor of 10 produced little change in the $^7$Be yield, while the factor of 100 did reduce the yield by ~30\%, but still not enough to achieve the required reduction factor of roughly 3. From our investigations, we believe that even the factor of 10 increase in this rate is unrealistically large.

\section{Non-thermal Neutrons in BBN}

In order to proceed with the analysis involving the non-thermal neutrons, it is essential to describe their energy spectrum. The 14.1-MeV neutrons produced by the $^3$H($d,n)^4$He reaction in BBN lose energy and eventually thermalize by scattering with other nuclei. If it is assumed that the time in the big bang is late enough that scattering from other neutrons can be ignored, the most important scattering target is $^1$H, which makes up over 90\% of the number density and is also the lightest isotope. The only other relevant nuclide, $^4$He, has a neutron scattering cross section  $\approx$ 30\% larger than $^1$H, but contributes less than 10\% of the total number density and is also less favorable for energy transfer due to its larger mass. Thus we only consider energy loss due to scattering from $^1$H. Our ultimate goal is to calculate the reactions induced by these neutrons before they thermalize.

Data for the $n-p$ total cross section are given in Ref.~\cite{Schwartz196936,Clement197251,larson}; we utilize these along with a power-law parameterization
\begin{equation}
\sigma(E)=\sigma_0(E/E_0)^\alpha,
\label{npsigma}
\end{equation}
where $E$ is the neutron energy and the constants are given by $\sigma_0= 0.685$ barns, $E_0 = 14.1 {\rm MeV}$, and $\alpha = - 0.834$. This parameterization reproduces the experimental data within 10\% for $3 < E < 20$ MeV. It overestimates the cross section for lower energies. However, we will use this parameterization in order to produce an upper limit on the effect of the non-thermal neutrons.

The energy transfer to the proton depends on the neutron scattering angle. Assuming that (1) the n-p differential cross section is isotropic in the center-of-mass system, (2) the neutron mass is equal to the proton mass, and (3) non-relativistic kinematics are valid, the energy loss of the neutron due to a single scattering event is uniformly distributed, from zero up to the initial neutron energy. These conditions are satisfied within a few percent for neutrons below 20 MeV; this assumption leads to significant simplifications in the analysis and will be utilized below.

We assumed that neutron thermalization occurs on a time scale much shorter than the local Hubble time and the timescale for changes in nuclear abundances.  Then the probability for neutrons of energy $E$ scattering from hydrogen in a time interval $\Delta t$ is given by
\begin{equation}
P(E)=\sigma(E) n_{\rm H} v\Delta t,
\label{prob1}
\end{equation}
where $\sigma (E)$ is the n-p total cross section, $n_{\rm H}$ is the $^1$H number density, $v$ is the neutron speed, and $\Delta t$ has been assumed to be sufficiently small that $P(E) << 1$. Likewise, the probability for neutrons of energy $E$ to scatter from hydrogen into the energy interval $[E^\prime, E^\prime + dE^\prime]$ is given by
\begin{equation}
P(E, E^\prime)dE^\prime = \sigma(E) n_{\rm H}  v\Delta t dE^\prime,
\label{prob2}
\end{equation}
 where we have made use of the uniform energy distribution discussed above.

The non-thermal neutron energy distribution consists of two contributions: a mono-energetic component of $n_0$ neutrons per unit volume accounting for 14.1-MeV neutrons which have not yet scattered, and a continuous distribution $f(E)$ describing the number of neutrons per unit volume per unit energy at lower energies. We define $E_0 = 14.1$ MeV.

$f(E)$ will evolve over the time interval $\Delta t$ as:
\begin{eqnarray}
\Delta f(E) &=& - f(E)P(E) + n_0 P(E_0, E)\label{fevolved}	
\\
&&+ \int_E^{E_0} f(E^\prime) P(E^\prime, E) dE^\prime.\nonumber 
\end{eqnarray}
The first term on the right hand side of Eq.~(\ref{fevolved}) describes the loss due to downscattering to lower energies, the second describes the increase due to the downscattering of 14.1-MeV neutrons, and the third describes the increase due to the downscattering of continuous neutrons at higher energies. With the assumption of a fast timescale for thermalization, $f(E)$ will be in dynamic equilibrium and $\Delta f(E) = 0$. Using these equations and $E = (1/2)mv^2$ gives
\begin{eqnarray}
- f(E) \sigma(E) E^{1/2} + n_0 \sigma(E) E^{-1/2} \\
+ \int _E^{E_0} f(E^\prime) \sigma(E^\prime) (E^\prime)^{-1/2} dE^\prime = 0.	\nonumber
\label{derived}
\end{eqnarray}
This implies that
\begin{equation}
f(E_0) = n_0/E_0.
\label{implied}
\end{equation}
If $\sigma(E)$ is given by Eq.~(\ref{npsigma}), Eq.~(\ref{derived}) can be solved analytically to yield the solution (for $\alpha \neq -1/2$),
\begin{equation}
f(E) ={{n_0}\over{E(\alpha + 1/2)}}\left[(E_o/E)^{\alpha+1/2} + \alpha - 1/2\right].
\end{equation}

\begin{figure}
\includegraphics[width=2.75in,angle=90]{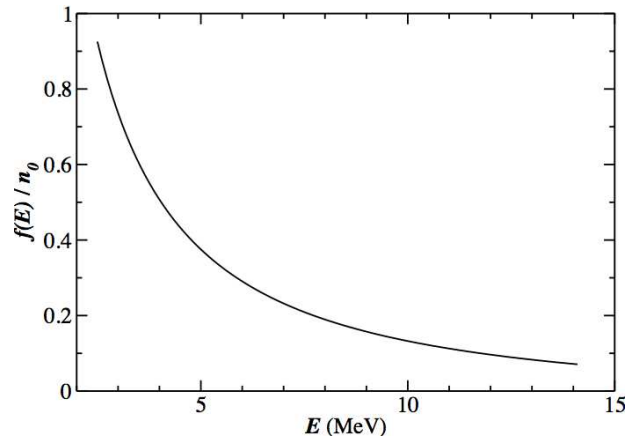}
\caption{Neutron energy distribution.}
\label{ndist}
\end{figure}

This distribution for $f(E)$ is shown in Fig.~\ref{ndist}. Because our model does not include any neutron sinks or upscattering, the distribution is highly singular at $E = 0$. The normalization down to some minimum energy can be defined:
\begin{eqnarray}
&N_m=\int_{E_m}^{E_0}f(E)dE={{n_0}\over{(\alpha+1/2)^2}}\\
&\times\left[(E_0/E_m)^{\alpha+1/2}-1+(\alpha^2-1/4)\log(E_0/E_m)\right].\nonumber
\label{normalization}
\end{eqnarray}
A minimum energy of $E_m$=2.5 MeV gives $N_m=2.975n_0$.

The number density of 14.1 MeV neutrons $n_0$ can be found by equating the rates for production and destruction. The destruction rate is given by:
\begin{equation}
r=n_0P(E)=n_0n_H\langle\sigma(E_0)v_0\rangle,
\label{destrate}
\end{equation}
where $E_0 = (1/2) mv^2$. The production rate is given by
\begin{equation}
r=n_tn_d\langle\sigma v\rangle_{td\rightarrow n\alpha}
\label{prodrate}
\end{equation}
and consequently $n_0$ can be calculated using
\begin{equation}
n_0=[n_tn_d\langle\sigma v\rangle_{td\rightarrow n\alpha}]/[n_H\sigma(E_0)v_0].
\end{equation}
Note that the quantities $n_H$, $n_d$, $n_t$ and $\langle\sigma v\rangle_{td\rightarrow n\alpha}$ are available in existing standard BBN calculations.

The reaction rate (reactions per unit volume per unit time) of the non-thermal neutrons with species $x$ is given by
\begin{equation}
r_x=n_xn_0\sigma_x(E_0)v_0+n_x\int_{E_m}^{E_0}f(E)\sigma_x(E)vdE,
\label{reacrate}
\end{equation}
where $\sigma_x(E)$ is the $n + X$ cross section. In this model the only temperature dependence is through $n_0$.  A temperature-independent reaction rate $\langle\sigma_x v\rangle$ can thus be defined as
\begin{equation}
\langle\sigma_x v\rangle=\langle\sigma_x(E_0)v_0\rangle +(1/n)\int_{E_m}^{E_0}f(E)\sigma_x(E)vdE
\label{tempindepen}
\end{equation}
which only needs to be calculated once.The rate of reactions defined by Eq.~(\ref{normalization}) can now be written
\begin{equation}
r_x=n_xn_0\langle\sigma_xv\rangle.
\label{eq18}
\end{equation}
With the above framework, it is now simple to implement the extra reaction flow due to non-thermal neutrons into the BBN calculations. 

\begin{figure}
\includegraphics[width=2.75in,angle=90]{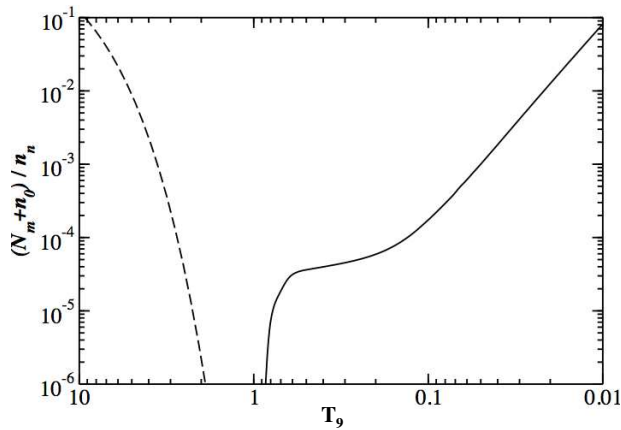}
\caption{The fraction of neutrons with energies above $E_m$= 2.5 MeV is shown by the solid curve. The fraction predicted by Maxwell-Boltzmann distribution is shown by the dashed curve.}
\label{fracneut}
\end{figure}

These show that the neutron energy distribution for high energy neutrons, that is, those above 2.5 MeV, is many orders of magnitude larger than the prediction of the Maxwell-Boltzmann distribution when the non-thermal contribution is included. Ê
See figures ~\ref{ndist} and \ref{fracneut}. ÊThe effects on the BBN abundances, Êhowever, were found to be at the $10^{-4}$ level or less, much smaller than the level of uncertainties resulting from other sources such as reaction rates. Details of several of the reactions considered are given below.

\begin{figure}
\includegraphics[width=2.75in,angle=90]{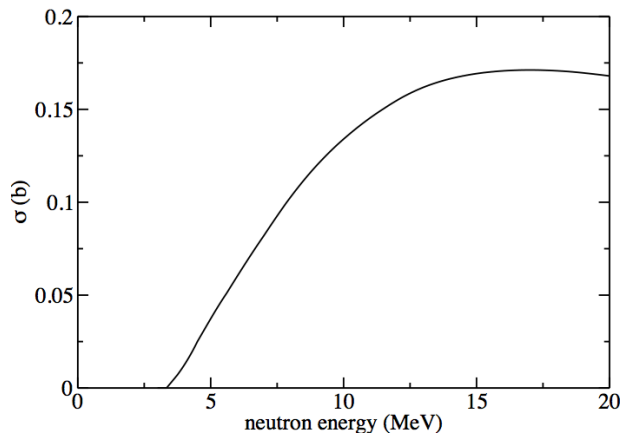}
\caption{The cross section for the $n + d \rightarrow 2n + p$ reaction as a function of incident neutron energy, from the ENDF/B-VI evaluation \cite{larson}.}
\label{nplusd}
\end{figure}

\subsubsection{\rm {$^2$H$(n,p)n$}}
Deuterium formation is crucial to all subsequent nucleosynthesis.  Non-thermal neutrons might allow the endothermic deuterium destruction $^2$H($n,p)n$ reaction to occur at an appreciable rate once nuclear statistical equilibrium for $n+p\rightleftharpoons ^2H+\gamma$ breaks down following alpha particle formation. The cross section for $^2$H($n,p)n$ is shown in Fig.~\ref{nplusd}. The inclusion of this reaction is found to increase the deuterium abundance by only 0.011\%, and the effects on other isotopes are an order of magnitude smaller. Note that this is an upper limit on the possible effects of the non-thermal neutrons on this reaction.

\subsubsection{\rm{$^3$He($n,p)^3$H}}
This reaction is exothermic, so can also proceed also via thermal neutrons on this reaction. Thermal neutrons are far more effective than non-thermal neutrons in producing reactions as a result of the $1/v$ law dependence of the neutron capture cross section. We therefore conclude that non-thermal effects will be negligible for this reaction.

\subsubsection{\rm {$^7$Be($n,$X)} reactions}
Neutron-induced reactions on $^7$Be producing the following exit channels were considered: $^2$H$+^6$Li, $n+p+^6$Li, $n+^3$He+$^4$He, and $^4$He+$^4$He. The first three are endothermic; the latter is exothermic, but highly suppressed for low-energy neutrons, as it cannot proceed via s-waves in the entrance channel. The cross sections for the reactions are not known. The maximum inelastic cross section for $n+^7$Li approaches 3 barns \cite{Evalnuc}. We have adopted this as an upper limit for the above reactions. Inclusion of $^7$Be($n, n ^3$He)$^4$He, then, was found to reduce the $^7$Be abundance by 0.015\%. Results for the other exit channels would be expected to be similar. Again, these are expected to be $upper\ limits$ on the effect of the non-thermal neutrons.

\section{Further Discussion of Thermal population of excited states and resonant enhancement}
The lowest-lying first excited states of the lightest nuclei are those of the mirrors $^7$Li and $^7$Be, at excitation energy 478 keV and 429 keV, respectively. ÊThese energies are tantalizingly close to BBN thermal energies, for example with alpha particle assembly at $kT\sim 100$ keV. ÊAccordingly, we have examined how potentially key nuclear reactions like $^7$Be($d,\gamma)^9$B, $^7$Be($d,p)2\alpha$, and $^7$Be($d,^3$He)$^6$Li could be altered if they were to proceed through a thermally populated first excited state, $^7$Be*. ÊWe have discussed this at length above (section II.10) for $^7$Be($d,^3$He)$^6$Li.  Some further remarks on possible thermal excitation effects are in order. 

Since the compound nucleus for $^2$H+$^7$Be, $^9$B, has a ground state that particle decays, and this reaction is 16 MeV up in excitation in the compound nucleus, it is unlikely that it would be strongly resonant. Without the sort of enhanced rate that would be produced by a resonance, Êit could only have a very small effect on BBN abundances. Thus, we did not include the $^7$Be($^2$H,$p)$ reaction in the BBN code.
Ê
The $^7$Be*($p,\gamma)^8$B reaction initiated from the thermally excited $^7$Be first excited state could produce an enhanced rate. However, it cannot have much of an effect on the BBN abundances. There is a 40 keV wide state in $^8$B, the compound nucleus, at 770 keV excitation, which from the $^7$Be first excited state, would be at 241 keV bombarding energy. However, the spins and parities of the entrance channel particles together with the spins and parities of either the ground state of $^8$B or this resonant state ($2^+$ and $1^+$ respectively) require a p-wave capture. Furthermore, this is a radiative capture, for which the cross section is generally roughly two orders of magnitude suppressed from that of a particle transfer reaction. When one imposes a Boltzmann factor to represent the abundance of the excited state, this reaction is further suppressed. ÊConsequently we did not include this reaction in the BBN code.

In the end, however, standard BBN yield alterations due to reactions through the $^7$Be* state are found to be small on account of its small thermal population.

\section{Conclusion}

We have examined a number of nuclear reactions likely to occur in BBN.  Some of these have not been included in BBN calculations before and some might be thought to figure in the $^{6,7}$Li problems.  We have discussed nuclear physics uncertainties and other issues ($e.g.$, non-thermal neutrons, thermal excitation of nuclear target states) and we have performed sensitivity studies with an extended BBN reaction network calculation of abundance yields to get an idea of what might be possible with these reactions.

In broad brush,  the reactions added to the BBN code had virtually no effect on the BBN abundances, although a few of them did have an effect when their rates were increased by the factor of x1000. In particular, the $^3$He($^3$H,$\gamma)^6$Li reaction did produce a large $^6$Li abundance enhancement, and the $^7$Be($^2$H,$^3$He)$^6$Li reaction produced a ~30\% decrease in the mass-7 abundance when their corresponding rates were increased by a factor of 1000. Although this multiplier is probably well outside the expected uncertainties for the rates for these reactions, these reactions might bear further study just to pin down the BBN abundances with the highest possible precision.

More important, however, is the conclusion from this study that the solutions to the problems of the $^7$Li and $^6$Li abundances are hard to find in the reaction rates. For $^6$Li, it may well be that the observed BBN abundances are simply incorrect; it has been suggested by Cayrel {\it et al.} \cite{2007A&A...473L..37C} that the claimed $^6$Li abundance is actually a spurious effect that is generated by convective Doppler shifts. However, the $^7$Li discrepancy is not as easy to ignore.  If both problems persist, however, their solution may well require nonstandard physics. Several papers have been published recently, some involving unstable particles in the early universe that could have affected BBN. One set of papers studied the effects of the reactions resulting from non-thermal decay products of the particles. These could destroy $^7$Li directly, reducing its abundance, and would make $^6$Li by producing non-thermal distributions of $^3$H and $^3$He from proton and neutron knockout reactions on $^4$He. These could then undergo $^4$He($^3$H,$n)^6$Li and $^4$He($^3$He,$p)^6$Li reactions respectively. However, as discussed by Kusakabe {\it et al.} \cite{Kusakabe:2008kf}, these solutions have the difficulty of also affecting the $^2$H and or $^3$He abundances, creating new problems. Note, however, that new cross sections for $^4$He($\gamma,p)^3$H and $^4$He($\gamma,n)^3$He may affect that conclusion \cite{0004-637X-680-2-846}. The second possible effect of unstable particles in the early universe would be if they were massive and negatively charged, so could attach to the nuclei as they were produced. A suggestion by Pospelov \cite{Pospelov:2006sc}, that such particles could enhance the $^4$He($d,\gamma)^6$Li reaction by many orders of magnitude by enabling a transfer reaction instead of the radiative capture is the critical feature of this solution. This has been found by several authors to solve both Li problems \cite{Pospelov:2006sc, Hamaguchi:2007mp, 1475-7516-2006-11-014, PhysRevD.78.083010, PhysRevD.76.063507, 0004-637X-680-2-846, Kawasaki2007436, Kawasaki:2008qe}. 

In summary, we find little effect on standard BBN abundance yields with the addition of the new nuclear physics we discuss here, even given a fair uncertainty in issues that bear on key reaction rates.  It could be argued, however, that the new nuclear reaction physics we examine here may yet be important for non-standard BBN scenarios with new particle physics.  Indeed, a conclusion of this paper is that there seems to be little chance of solving either of the \lq\lq lithium problems\rq\rq\ by conventional nuclear physics means and, if these problems stand up to future observations, we may be forced into just such non-standard BBN scenarios.

\begin{acknowledgments}
The authors acknowledge early contributions to this project by T. Luu and insightful discussions with Wick Haxton and Janilee Benitez.  This work was performed under the auspices of the Lawrence Livermore National Security, LLC, (LLNS) under Contract No. DE-AC52-07NA27344, and DOE grants DE-FG02-88ER40387 and DE-FG52-09NA29455. CRB acknowledges support from LLNL LDRD grant ER-066.  GMF acknowledges support from NSF Grant No. PHY-06-53626 at UCSD and CJS would like to thank ASU for support.  This paper is number LLNL-JRNL-446832.
\end{acknowledgments}

\bibliography{mybiblio}

\begin{thebibliography}{70}
\expandafter\ifx\csname natexlab\endcsname\relax\def\natexlab#1{#1}\fi
\expandafter\ifx\csname bibnamefont\endcsname\relax
  \def\bibnamefont#1{#1}\fi
\expandafter\ifx\csname bibfnamefont\endcsname\relax
  \def\bibfnamefont#1{#1}\fi
\expandafter\ifx\csname citenamefont\endcsname\relax
  \def\citenamefont#1{#1}\fi
\expandafter\ifx\csname url\endcsname\relax
  \def\url#1{\texttt{#1}}\fi
\expandafter\ifx\csname urlprefix\endcsname\relax\def\urlprefix{URL }\fi
\providecommand{\bibinfo}[2]{#2}
\providecommand{\eprint}[2][]{\url{#2}}

\bibitem[{\citenamefont{Wagoner}(1969)}]{wag69}
\bibinfo{author}{\bibfnamefont{R.~V.} \bibnamefont{Wagoner}},
  \bibinfo{journal}{Ann. Rev. Astron. Astrophys.} \textbf{\bibinfo{volume}{7}},
  \bibinfo{pages}{553} (\bibinfo{year}{1969}).

\bibitem[{\citenamefont{Wagoner et~al.}(1967)\citenamefont{Wagoner, Fowler, and
  Hoyle}}]{wfh}
\bibinfo{author}{\bibfnamefont{R.~V.} \bibnamefont{Wagoner}},
  \bibinfo{author}{\bibfnamefont{W.~A.} \bibnamefont{Fowler}},
  \bibnamefont{and} \bibinfo{author}{\bibfnamefont{F.}~\bibnamefont{Hoyle}},
  \bibinfo{journal}{Astrophys.\ J.} \textbf{\bibinfo{volume}{148}},
  \bibinfo{pages}{3} (\bibinfo{year}{1967}).

\bibitem[{\citenamefont{Yang et~al.}(1984)\citenamefont{Yang, Turner, Steigman,
  Schramm, and Olive}}]{JYang}
\bibinfo{author}{\bibfnamefont{J.-M.} \bibnamefont{Yang}},
  \bibinfo{author}{\bibfnamefont{M.~S.} \bibnamefont{Turner}},
  \bibinfo{author}{\bibfnamefont{G.}~\bibnamefont{Steigman}},
  \bibinfo{author}{\bibfnamefont{D.~N.} \bibnamefont{Schramm}},
  \bibnamefont{and} \bibinfo{author}{\bibfnamefont{K.~A.} \bibnamefont{Olive}},
  \bibinfo{journal}{Astrophys J.} \textbf{\bibinfo{volume}{281}},
  \bibinfo{pages}{493} (\bibinfo{year}{1984}).

\bibitem[{\citenamefont{Walker et~al.}(1991)\citenamefont{Walker, Steigman,
  Schramm, Olive, and Kang}}]{Walker:1991ap}
\bibinfo{author}{\bibfnamefont{T.~P.} \bibnamefont{Walker}},
  \bibinfo{author}{\bibfnamefont{G.}~\bibnamefont{Steigman}},
  \bibinfo{author}{\bibfnamefont{D.~N.} \bibnamefont{Schramm}},
  \bibinfo{author}{\bibfnamefont{K.~A.} \bibnamefont{Olive}}, \bibnamefont{and}
  \bibinfo{author}{\bibfnamefont{H.-S.} \bibnamefont{Kang}},
  \bibinfo{journal}{Astrophys. J.} \textbf{\bibinfo{volume}{376}},
  \bibinfo{pages}{51} (\bibinfo{year}{1991}).

\bibitem[{\citenamefont{{Krauss} and {Romanelli}}(1990)}]{1990ApJ...358...47K}
\bibinfo{author}{\bibfnamefont{L.~M.} \bibnamefont{{Krauss}}} \bibnamefont{and}
  \bibinfo{author}{\bibfnamefont{P.}~\bibnamefont{{Romanelli}}},
  \bibinfo{journal}{\apj} \textbf{\bibinfo{volume}{358}}, \bibinfo{pages}{47}
  (\bibinfo{year}{1990}).

\bibitem[{\citenamefont{Smith et~al.}(1993)\citenamefont{Smith, Kawano, and
  Malaney}}]{skm}
\bibinfo{author}{\bibfnamefont{M.~S.} \bibnamefont{Smith}},
  \bibinfo{author}{\bibfnamefont{L.~H.} \bibnamefont{Kawano}},
  \bibnamefont{and} \bibinfo{author}{\bibfnamefont{R.~A.}
  \bibnamefont{Malaney}}, \bibinfo{journal}{Astrophys.\ J.\ Suppl.}
  \textbf{\bibinfo{volume}{85}}, \bibinfo{pages}{219} (\bibinfo{year}{1993}).

\bibitem[{\citenamefont{Kawano}(1992)}]{kawano}
\bibinfo{author}{\bibfnamefont{L.}~\bibnamefont{Kawano}},
  \bibinfo{journal}{NASA STI/Recon Technical Report N}
  \textbf{\bibinfo{volume}{92}}, \bibinfo{pages}{25163} (\bibinfo{year}{1992}).

\bibitem[{\citenamefont{Kawano}(1988)}]{kawano1}
\bibinfo{author}{\bibfnamefont{L.}~\bibnamefont{Kawano}}
  (\bibinfo{year}{1988}), \eprint{FERMILAB-PUB-88/34-A}.

\bibitem[{\citenamefont{Nollett and Burles}(2000)}]{Nollett:2000fh}
\bibinfo{author}{\bibfnamefont{K.~M.} \bibnamefont{Nollett}} \bibnamefont{and}
  \bibinfo{author}{\bibfnamefont{S.}~\bibnamefont{Burles}},
  \bibinfo{journal}{Phys. Rev.} \textbf{\bibinfo{volume}{D61}},
  \bibinfo{pages}{123505} (\bibinfo{year}{2000}), \eprint{astro-ph/0001440}.

\bibitem[{\citenamefont{Schramm and Turner}(1998)}]{Schramm:1997vs}
\bibinfo{author}{\bibfnamefont{D.~N.} \bibnamefont{Schramm}} \bibnamefont{and}
  \bibinfo{author}{\bibfnamefont{M.~S.} \bibnamefont{Turner}},
  \bibinfo{journal}{Rev. Mod. Phys.} \textbf{\bibinfo{volume}{70}},
  \bibinfo{pages}{303} (\bibinfo{year}{1998}), \eprint{astro-ph/9706069}.

\bibitem[{\citenamefont{Smith et~al.}(2009)\citenamefont{Smith, Fuller, and
  Smith}}]{ourcode}
\bibinfo{author}{\bibfnamefont{C.~J.} \bibnamefont{Smith}},
  \bibinfo{author}{\bibfnamefont{G.~M.} \bibnamefont{Fuller}},
  \bibnamefont{and} \bibinfo{author}{\bibfnamefont{M.~S.} \bibnamefont{Smith}},
  \bibinfo{journal}{Physical Review D (Particles, Fields, Gravitation, and
  Cosmology)} \textbf{\bibinfo{volume}{79}}, \bibinfo{eid}{105001}
  (pages~\bibinfo{numpages}{10}) (\bibinfo{year}{2009}),
  \urlprefix\url{http://link.aps.org/abstract/PRD/v79/e105001}.

\bibitem[{\citenamefont{Smith and Fuller}(2010)}]{coulfac}
\bibinfo{author}{\bibfnamefont{C.~J.} \bibnamefont{Smith}} \bibnamefont{and}
  \bibinfo{author}{\bibfnamefont{G.~M.} \bibnamefont{Fuller}},
  \bibinfo{journal}{Phys. Rev.} \textbf{\bibinfo{volume}{D81}},
  \bibinfo{pages}{065027} (\bibinfo{year}{2010}), \eprint{0905.2781}.

\bibitem[{\citenamefont{Olive et~al.}(1997)\citenamefont{Olive, Steigman, and
  Skillman}}]{Olive}
\bibinfo{author}{\bibfnamefont{K.~A.} \bibnamefont{Olive}},
  \bibinfo{author}{\bibfnamefont{G.}~\bibnamefont{Steigman}}, \bibnamefont{and}
  \bibinfo{author}{\bibfnamefont{E.~D.} \bibnamefont{Skillman}},
  \bibinfo{journal}{Astrophys.\ J.} \textbf{\bibinfo{volume}{483}},
  \bibinfo{pages}{788} (\bibinfo{year}{1997}).

\bibitem[{\citenamefont{Steigman}(2006)}]{Steigman1}
\bibinfo{author}{\bibfnamefont{G.}~\bibnamefont{Steigman}},
  \bibinfo{journal}{Int.\ J.\ Mod.\ Phys.\ E} \textbf{\bibinfo{volume}{15}},
  \bibinfo{pages}{1} (\bibinfo{year}{2006}).

\bibitem[{\citenamefont{{Steigman}}(2007)}]{steig}
\bibinfo{author}{\bibfnamefont{G.}~\bibnamefont{{Steigman}}},
  \bibinfo{journal}{Annual Review of Nuclear and Particle Science}
  \textbf{\bibinfo{volume}{57}}, \bibinfo{pages}{463} (\bibinfo{year}{2007}),
  \eprint{0712.1100}.

\bibitem[{\citenamefont{{Asplund} et~al.}(2006)\citenamefont{{Asplund},
  {Lambert}, {Nissen}, {Primas}, and {Smith}}}]{asplund}
\bibinfo{author}{\bibfnamefont{M.}~\bibnamefont{{Asplund}}},
  \bibinfo{author}{\bibfnamefont{D.~L.} \bibnamefont{{Lambert}}},
  \bibinfo{author}{\bibfnamefont{P.~E.} \bibnamefont{{Nissen}}},
  \bibinfo{author}{\bibfnamefont{F.}~\bibnamefont{{Primas}}}, \bibnamefont{and}
  \bibinfo{author}{\bibfnamefont{V.~V.} \bibnamefont{{Smith}}},
  \bibinfo{journal}{\apj} \textbf{\bibinfo{volume}{644}}, \bibinfo{pages}{229}
  (\bibinfo{year}{2006}), \eprint{arXiv:astro-ph/0510636}.

\bibitem[{\citenamefont{Tegmark et~al.}(2004)}]{WMAP}
\bibinfo{author}{\bibfnamefont{M.}~\bibnamefont{Tegmark}} \bibnamefont{et~al.}
  (\bibinfo{collaboration}{SDSS Collaboration}), \bibinfo{journal}{Phys.\ Rev.\
  D} \textbf{\bibinfo{volume}{69}}, \bibinfo{pages}{103501}
  (\bibinfo{year}{2004}).

\bibitem[{\citenamefont{Spergel et~al.}(2003)}]{WMAP1}
\bibinfo{author}{\bibfnamefont{D.~N.} \bibnamefont{Spergel}}
  \bibnamefont{et~al.}, \bibinfo{journal}{Astrophys.\ J.\ Suppl.}
  \textbf{\bibinfo{volume}{148}}, \bibinfo{pages}{175} (\bibinfo{year}{2003}).

\bibitem[{\citenamefont{Spergel et~al.}(2006)}]{3yrwmap}
\bibinfo{author}{\bibfnamefont{D.~N.} \bibnamefont{Spergel}}
  \bibnamefont{et~al.} (\bibinfo{year}{2006}), \eprint{astro-ph/0603449}.

\bibitem[{\citenamefont{Bennett et~al.}(2003)}]{Bennett:2003ca}
\bibinfo{author}{\bibfnamefont{C.}~\bibnamefont{Bennett}} \bibnamefont{et~al.}
  (\bibinfo{collaboration}{WMAP}), \bibinfo{journal}{Astrophys. J. Suppl.}
  \textbf{\bibinfo{volume}{148}}, \bibinfo{pages}{97} (\bibinfo{year}{2003}),
  \eprint{astro-ph/0302208}.

\bibitem[{\citenamefont{Komatsu et~al.}(2010)}]{Komatsu:2010fb}
\bibinfo{author}{\bibfnamefont{E.}~\bibnamefont{Komatsu}} \bibnamefont{et~al.}
  (\bibinfo{year}{2010}), \eprint{1001.4538}.

\bibitem[{\citenamefont{Kirkman et~al.}(2003)\citenamefont{Kirkman, Tytler,
  Suzuki, O'Meara, , and Lubin}}]{Tytler}
\bibinfo{author}{\bibfnamefont{D.}~\bibnamefont{Kirkman}},
  \bibinfo{author}{\bibfnamefont{D.}~\bibnamefont{Tytler}},
  \bibinfo{author}{\bibfnamefont{N.}~\bibnamefont{Suzuki}},
  \bibinfo{author}{\bibfnamefont{J.~M.} \bibnamefont{O'Meara}}, ,
  \bibnamefont{and} \bibinfo{author}{\bibfnamefont{D.}~\bibnamefont{Lubin}},
  \bibinfo{journal}{Astrophys.\ J.\ Suppl.} \textbf{\bibinfo{volume}{149}},
  \bibinfo{pages}{1} (\bibinfo{year}{2003}).

\bibitem[{\citenamefont{O'Meara et~al.}(2001)\citenamefont{O'Meara, Tytler,
  Kirkman, Suzuki, Prochaska, Lubin, and Wolfe}}]{Omeara}
\bibinfo{author}{\bibfnamefont{J.~M.} \bibnamefont{O'Meara}},
  \bibinfo{author}{\bibfnamefont{D.}~\bibnamefont{Tytler}},
  \bibinfo{author}{\bibfnamefont{D.}~\bibnamefont{Kirkman}},
  \bibinfo{author}{\bibfnamefont{N.}~\bibnamefont{Suzuki}},
  \bibinfo{author}{\bibfnamefont{J.~X.} \bibnamefont{Prochaska}},
  \bibinfo{author}{\bibfnamefont{D.}~\bibnamefont{Lubin}}, \bibnamefont{and}
  \bibinfo{author}{\bibfnamefont{A.~M.} \bibnamefont{Wolfe}},
  \bibinfo{journal}{Astrophys.\ J.} \textbf{\bibinfo{volume}{552}},
  \bibinfo{pages}{718} (\bibinfo{year}{2001}).

\bibitem[{pla()}]{planck}
\urlprefix\url{http://www.rssd.esa.int/index.php?project=planck}.

\bibitem[{\citenamefont{Spite and Spite}(1982)}]{spite}
\bibinfo{author}{\bibfnamefont{F.}~\bibnamefont{Spite}} \bibnamefont{and}
  \bibinfo{author}{\bibfnamefont{M.}~\bibnamefont{Spite}},
  \bibinfo{journal}{Astron.\ Astrophys.} \textbf{\bibinfo{volume}{115}},
  \bibinfo{pages}{357} (\bibinfo{year}{1982}).

\bibitem[{\citenamefont{{Ryan} et~al.}(1999)\citenamefont{{Ryan}, {Norris}, and
  {Beers}}}]{1999ApJ...523..654R}
\bibinfo{author}{\bibfnamefont{S.~G.} \bibnamefont{{Ryan}}},
  \bibinfo{author}{\bibfnamefont{J.~E.} \bibnamefont{{Norris}}},
  \bibnamefont{and} \bibinfo{author}{\bibfnamefont{T.~C.}
  \bibnamefont{{Beers}}}, \bibinfo{journal}{\apj}
  \textbf{\bibinfo{volume}{523}}, \bibinfo{pages}{654} (\bibinfo{year}{1999}),
  \eprint{arXiv:astro-ph/9903059}.

\bibitem[{\citenamefont{{Cayrel} et~al.}(2007)\citenamefont{{Cayrel},
  {Steffen}, {Chand}, {Bonifacio}, {Spite}, {Spite}, {Petitjean}, {Ludwig}, and
  {Caffau}}}]{2007A&A...473L..37C}
\bibinfo{author}{\bibfnamefont{R.}~\bibnamefont{{Cayrel}}},
  \bibinfo{author}{\bibfnamefont{M.}~\bibnamefont{{Steffen}}},
  \bibinfo{author}{\bibfnamefont{H.}~\bibnamefont{{Chand}}},
  \bibinfo{author}{\bibfnamefont{P.}~\bibnamefont{{Bonifacio}}},
  \bibinfo{author}{\bibfnamefont{M.}~\bibnamefont{{Spite}}},
  \bibinfo{author}{\bibfnamefont{F.}~\bibnamefont{{Spite}}},
  \bibinfo{author}{\bibfnamefont{P.}~\bibnamefont{{Petitjean}}},
  \bibinfo{author}{\bibfnamefont{H.}~\bibnamefont{{Ludwig}}}, \bibnamefont{and}
  \bibinfo{author}{\bibfnamefont{E.}~\bibnamefont{{Caffau}}},
  \bibinfo{journal}{Astron. Astrophys.} \textbf{\bibinfo{volume}{473}},
  \bibinfo{pages}{L37} (\bibinfo{year}{2007}), \eprint{0708.3819}.

\bibitem[{\citenamefont{Cyburt et~al.}(2006)\citenamefont{Cyburt, Ellis,
  Fields, Olive, and Spanos}}]{1475-7516-2006-11-014}
\bibinfo{author}{\bibfnamefont{R.~H.} \bibnamefont{Cyburt}},
  \bibinfo{author}{\bibfnamefont{J.}~\bibnamefont{Ellis}},
  \bibinfo{author}{\bibfnamefont{B.~D.} \bibnamefont{Fields}},
  \bibinfo{author}{\bibfnamefont{K.~A.} \bibnamefont{Olive}}, \bibnamefont{and}
  \bibinfo{author}{\bibfnamefont{V.~C.} \bibnamefont{Spanos}},
  \bibinfo{journal}{Journal of Cosmology and Astroparticle Physics}
  \textbf{\bibinfo{volume}{2006}}, \bibinfo{pages}{014} (\bibinfo{year}{2006}),
  \urlprefix\url{http://stacks.iop.org/1475-7516/2006/i=11/a=014}.

\bibitem[{\citenamefont{Jedamzik}(2000)}]{Jedamzik:1999di}
\bibinfo{author}{\bibfnamefont{K.}~\bibnamefont{Jedamzik}},
  \bibinfo{journal}{Phys. Rev. Lett.} \textbf{\bibinfo{volume}{84}},
  \bibinfo{pages}{3248} (\bibinfo{year}{2000}), \eprint{astro-ph/9909445}.

\bibitem[{\citenamefont{{Dimopoulos} et~al.}(1988)\citenamefont{{Dimopoulos},
  {Esmailzadeh}, {Hall}, and {Starkman}}}]{1988PhRvL..60Q...7D}
\bibinfo{author}{\bibfnamefont{S.}~\bibnamefont{{Dimopoulos}}},
  \bibinfo{author}{\bibfnamefont{R.}~\bibnamefont{{Esmailzadeh}}},
  \bibinfo{author}{\bibfnamefont{L.~J.} \bibnamefont{{Hall}}},
  \bibnamefont{and} \bibinfo{author}{\bibfnamefont{G.~D.}
  \bibnamefont{{Starkman}}}, \bibinfo{journal}{Physical Review Letters}
  \textbf{\bibinfo{volume}{60}}, \bibinfo{pages}{7} (\bibinfo{year}{1988}).

\bibitem[{\citenamefont{Nollett et~al.}(1997)\citenamefont{Nollett, Lemoine,
  and Schramm}}]{Nollett:1996ef}
\bibinfo{author}{\bibfnamefont{K.~M.} \bibnamefont{Nollett}},
  \bibinfo{author}{\bibfnamefont{M.}~\bibnamefont{Lemoine}}, \bibnamefont{and}
  \bibinfo{author}{\bibfnamefont{D.~N.} \bibnamefont{Schramm}},
  \bibinfo{journal}{Phys. Rev.} \textbf{\bibinfo{volume}{C56}},
  \bibinfo{pages}{1144} (\bibinfo{year}{1997}), \eprint{astro-ph/9612197}.

\bibitem[{\citenamefont{Cyburt and Pospelov}(2009)}]{Cyburt:2009cf}
\bibinfo{author}{\bibfnamefont{R.~H.} \bibnamefont{Cyburt}} \bibnamefont{and}
  \bibinfo{author}{\bibfnamefont{M.}~\bibnamefont{Pospelov}}
  (\bibinfo{year}{2009}), \eprint{astro-ph.CO/0906.4373}.

\bibitem[{\citenamefont{Smith et~al.}(2006)\citenamefont{Smith, Fuller,
  Kishimoto, and Abazajian}}]{sfka}
\bibinfo{author}{\bibfnamefont{C.~J.} \bibnamefont{Smith}},
  \bibinfo{author}{\bibfnamefont{G.~M.} \bibnamefont{Fuller}},
  \bibinfo{author}{\bibfnamefont{C.~T.} \bibnamefont{Kishimoto}},
  \bibnamefont{and} \bibinfo{author}{\bibfnamefont{K.~N.}
  \bibnamefont{Abazajian}}, \bibinfo{journal}{Phys. Rev.}
  \textbf{\bibinfo{volume}{D74}}, \bibinfo{pages}{085008}
  (\bibinfo{year}{2006}), \eprint{astro-ph/0608377}.

\bibitem[{\citenamefont{Abazajian et~al.}(2005)\citenamefont{Abazajian, Bell,
  Fuller, and Wong}}]{abfw}
\bibinfo{author}{\bibfnamefont{K.}~\bibnamefont{Abazajian}},
  \bibinfo{author}{\bibfnamefont{N.~F.} \bibnamefont{Bell}},
  \bibinfo{author}{\bibfnamefont{G.~M.} \bibnamefont{Fuller}},
  \bibnamefont{and} \bibinfo{author}{\bibfnamefont{Y.~Y.~Y.}
  \bibnamefont{Wong}}, \bibinfo{journal}{Phys.\ Rev.\ D}
  \textbf{\bibinfo{volume}{72}}, \bibinfo{pages}{063004}
  (\bibinfo{year}{2005}).

\bibitem[{\citenamefont{Burles et~al.}(2001)\citenamefont{Burles, Nollett, and
  Turner}}]{Burles:2000zk}
\bibinfo{author}{\bibfnamefont{S.}~\bibnamefont{Burles}},
  \bibinfo{author}{\bibfnamefont{K.~M.} \bibnamefont{Nollett}},
  \bibnamefont{and} \bibinfo{author}{\bibfnamefont{M.~S.}
  \bibnamefont{Turner}}, \bibinfo{journal}{Astrophys. J.}
  \textbf{\bibinfo{volume}{552}}, \bibinfo{pages}{L1} (\bibinfo{year}{2001}),
  \eprint{astro-ph/0010171}.

\bibitem[{\citenamefont{Steigman}(2007)}]{Steigman:2007xt}
\bibinfo{author}{\bibfnamefont{G.}~\bibnamefont{Steigman}},
  \bibinfo{journal}{Ann. Rev. Nucl. Part. Sci.} \textbf{\bibinfo{volume}{57}},
  \bibinfo{pages}{463} (\bibinfo{year}{2007}), \eprint{0712.1100}.

\bibitem[{\citenamefont{Voronchev et~al.}(2008)\citenamefont{Voronchev, Nakao,
  and Nakamura}}]{Voronchev:2008zz}
\bibinfo{author}{\bibfnamefont{V.~T.} \bibnamefont{Voronchev}},
  \bibinfo{author}{\bibfnamefont{Y.}~\bibnamefont{Nakao}}, \bibnamefont{and}
  \bibinfo{author}{\bibfnamefont{M.}~\bibnamefont{Nakamura}},
  \bibinfo{journal}{J. Cosmol. Astropart. Phys.}
  \textbf{\bibinfo{volume}{0805}}, \bibinfo{pages}{010} (\bibinfo{year}{2008}).

\bibitem[{\citenamefont{Boyd}(2008)}]{boydbook}
\bibinfo{author}{\bibfnamefont{R.~N.} \bibnamefont{Boyd}},
  \emph{\bibinfo{title}{An Introduction to Nuclear Astrophysics}}
  (\bibinfo{publisher}{U. Chicago}, \bibinfo{year}{2008}),
  \bibinfo{edition}{1st} ed.

\bibitem[{\citenamefont{Brune et~al.}(1991)\citenamefont{Brune, Kavanagh,
  Kellogg, and Wang}}]{Brune:1991zza}
\bibinfo{author}{\bibfnamefont{C.~R.} \bibnamefont{Brune}},
  \bibinfo{author}{\bibfnamefont{R.~W.} \bibnamefont{Kavanagh}},
  \bibinfo{author}{\bibfnamefont{S.~E.} \bibnamefont{Kellogg}},
  \bibnamefont{and} \bibinfo{author}{\bibfnamefont{T.~R.} \bibnamefont{Wang}},
  \bibinfo{journal}{Phys. Rev.} \textbf{\bibinfo{volume}{C43}},
  \bibinfo{pages}{875} (\bibinfo{year}{1991}).

\bibitem[{\citenamefont{Rath et~al.}(1990)\citenamefont{Rath, Boyd, Hausman,
  Islam, and Kolnicki}}]{Rath1990338}
\bibinfo{author}{\bibfnamefont{D.~P.} \bibnamefont{Rath}},
  \bibinfo{author}{\bibfnamefont{R.~N.} \bibnamefont{Boyd}},
  \bibinfo{author}{\bibfnamefont{H.~J.} \bibnamefont{Hausman}},
  \bibinfo{author}{\bibfnamefont{M.~S.} \bibnamefont{Islam}}, \bibnamefont{and}
  \bibinfo{author}{\bibfnamefont{G.~W.} \bibnamefont{Kolnicki}},
  \bibinfo{journal}{Nuclear Physics A} \textbf{\bibinfo{volume}{515}},
  \bibinfo{pages}{338 } (\bibinfo{year}{1990}), ISSN \bibinfo{issn}{0375-9474}.

\bibitem[{\citenamefont{Yan}(1995)}]{jyanphd}
\bibinfo{author}{\bibfnamefont{J.}~\bibnamefont{Yan}}, \bibinfo{journal}{Ph.D
  Thesis, Colorado School of Mines}  (\bibinfo{year}{1995}).

\bibitem[{\citenamefont{Duggan et~al.}(1963)\citenamefont{Duggan, Miller, and
  Gabbard}}]{Duggan1963336}
\bibinfo{author}{\bibfnamefont{J.}~\bibnamefont{Duggan}},
  \bibinfo{author}{\bibfnamefont{P.}~\bibnamefont{Miller}}, \bibnamefont{and}
  \bibinfo{author}{\bibfnamefont{R.}~\bibnamefont{Gabbard}},
  \bibinfo{journal}{Nuclear Physics} \textbf{\bibinfo{volume}{46}},
  \bibinfo{pages}{336 } (\bibinfo{year}{1963}), ISSN \bibinfo{issn}{0029-5582}.

\bibitem[{\citenamefont{Pallone}(2000)}]{pallone}
\bibinfo{author}{\bibfnamefont{A.~K.} \bibnamefont{Pallone}},
  \bibinfo{journal}{Ph.D Thesis, Colorado School of Mines}
  (\bibinfo{year}{2000}).

\bibitem[{\citenamefont{Warner et~al.}(1987)\citenamefont{Warner, Vaughan,
  Ditusa, Rovine, Wakeland, Browne, Darden, Sen, Nadasen, Basak
  et~al.}}]{Warner1987339}
\bibinfo{author}{\bibfnamefont{R.~E.} \bibnamefont{Warner}},
  \bibinfo{author}{\bibfnamefont{B.~A.} \bibnamefont{Vaughan}},
  \bibinfo{author}{\bibfnamefont{J.~A.} \bibnamefont{Ditusa}},
  \bibinfo{author}{\bibfnamefont{J.~W.} \bibnamefont{Rovine}},
  \bibinfo{author}{\bibfnamefont{R.~S.} \bibnamefont{Wakeland}},
  \bibinfo{author}{\bibfnamefont{C.~P.} \bibnamefont{Browne}},
  \bibinfo{author}{\bibfnamefont{S.~E.} \bibnamefont{Darden}},
  \bibinfo{author}{\bibfnamefont{S.}~\bibnamefont{Sen}},
  \bibinfo{author}{\bibfnamefont{A.}~\bibnamefont{Nadasen}},
  \bibinfo{author}{\bibfnamefont{A.}~\bibnamefont{Basak}},
  \bibnamefont{et~al.}, \bibinfo{journal}{Nuclear Physics A}
  \textbf{\bibinfo{volume}{470}}, \bibinfo{pages}{339 } (\bibinfo{year}{1987}),
  ISSN \bibinfo{issn}{0375-9474}.

\bibitem[{\citenamefont{Forsyth and Perry}(1965)}]{Forsyth1965517}
\bibinfo{author}{\bibfnamefont{P.~D.} \bibnamefont{Forsyth}} \bibnamefont{and}
  \bibinfo{author}{\bibfnamefont{R.~R.} \bibnamefont{Perry}},
  \bibinfo{journal}{Nuclear Physics} \textbf{\bibinfo{volume}{67}},
  \bibinfo{pages}{517 } (\bibinfo{year}{1965}), ISSN \bibinfo{issn}{0029-5582}.

\bibitem[{\citenamefont{Cecil et~al.}(1983)\citenamefont{Cecil, Fahlsing,
  Jarmie, Hardekopf, and Martinez}}]{PhysRevC.27.6}
\bibinfo{author}{\bibfnamefont{F.~E.} \bibnamefont{Cecil}},
  \bibinfo{author}{\bibfnamefont{R.~F.} \bibnamefont{Fahlsing}},
  \bibinfo{author}{\bibfnamefont{N.}~\bibnamefont{Jarmie}},
  \bibinfo{author}{\bibfnamefont{R.~A.} \bibnamefont{Hardekopf}},
  \bibnamefont{and} \bibinfo{author}{\bibfnamefont{R.}~\bibnamefont{Martinez}},
  \bibinfo{journal}{Phys. Rev. C} \textbf{\bibinfo{volume}{27}},
  \bibinfo{pages}{6} (\bibinfo{year}{1983}).

\bibitem[{\citenamefont{Young et~al.}(1970)\citenamefont{Young, Blatt, and
  Seyler}}]{PhysRevLett.25.1764}
\bibinfo{author}{\bibfnamefont{A.~M.} \bibnamefont{Young}},
  \bibinfo{author}{\bibfnamefont{S.~L.} \bibnamefont{Blatt}}, \bibnamefont{and}
  \bibinfo{author}{\bibfnamefont{R.~G.} \bibnamefont{Seyler}},
  \bibinfo{journal}{Phys. Rev. Lett.} \textbf{\bibinfo{volume}{25}},
  \bibinfo{pages}{1764} (\bibinfo{year}{1970}).

\bibitem[{\citenamefont{Blatt et~al.}(1968)\citenamefont{Blatt, Young, Ling,
  Moon, and Porterfield}}]{Blatt:1968zz}
\bibinfo{author}{\bibfnamefont{S.~L.} \bibnamefont{Blatt}},
  \bibinfo{author}{\bibfnamefont{A.~M.} \bibnamefont{Young}},
  \bibinfo{author}{\bibfnamefont{S.~C.} \bibnamefont{Ling}},
  \bibinfo{author}{\bibfnamefont{K.~J.} \bibnamefont{Moon}}, \bibnamefont{and}
  \bibinfo{author}{\bibfnamefont{C.~D.} \bibnamefont{Porterfield}},
  \bibinfo{journal}{Phys. Rev.} \textbf{\bibinfo{volume}{176}},
  \bibinfo{pages}{1147} (\bibinfo{year}{1968}).

\bibitem[{\citenamefont{website}(2010)}]{NACRE}
\bibinfo{author}{\bibfnamefont{N.~A. C. R.~E.} \bibnamefont{website}}
  (\bibinfo{year}{2010}),
  \urlprefix\url{http://pntpm.ulb.ac.be/Nacre/nacre_d.htm}.

\bibitem[{\citenamefont{Serpico et~al.}(2004)}]{Serpico:2004gx}
\bibinfo{author}{\bibfnamefont{P.~D.} \bibnamefont{Serpico}}
  \bibnamefont{et~al.}, \bibinfo{journal}{J. Cosmol. Astropart. Phys.}
  \textbf{\bibinfo{volume}{0412}}, \bibinfo{pages}{010} (\bibinfo{year}{2004}),
  \eprint{astro-ph/0408076}.

\bibitem[{\citenamefont{Fukugita and Kajino}(1990)}]{Fukugita:1990kj}
\bibinfo{author}{\bibfnamefont{M.}~\bibnamefont{Fukugita}} \bibnamefont{and}
  \bibinfo{author}{\bibfnamefont{T.}~\bibnamefont{Kajino}},
  \bibinfo{journal}{Phys. Rev.} \textbf{\bibinfo{volume}{D42}},
  \bibinfo{pages}{4251} (\bibinfo{year}{1990}).

\bibitem[{\citenamefont{Madsen}(1990)}]{PhysRevD.41.2472}
\bibinfo{author}{\bibfnamefont{J.}~\bibnamefont{Madsen}},
  \bibinfo{journal}{Phys. Rev. D} \textbf{\bibinfo{volume}{41}},
  \bibinfo{pages}{2472} (\bibinfo{year}{1990}).

\bibitem[{\citenamefont{{Boyd} and {Kajino}}(1989)}]{1989ApJ...336L..55B}
\bibinfo{author}{\bibfnamefont{R.~N.} \bibnamefont{{Boyd}}} \bibnamefont{and}
  \bibinfo{author}{\bibfnamefont{T.}~\bibnamefont{{Kajino}}},
  \bibinfo{journal}{Astrophys. J. L.} \textbf{\bibinfo{volume}{336}},
  \bibinfo{pages}{L55} (\bibinfo{year}{1989}).

\bibitem[{\citenamefont{{Taylor} and {Phillips}}(1969)}]{1969NuPhA.126..615T}
\bibinfo{author}{\bibfnamefont{M.~C.} \bibnamefont{{Taylor}}} \bibnamefont{and}
  \bibinfo{author}{\bibfnamefont{G.~C.} \bibnamefont{{Phillips}}},
  \bibinfo{journal}{Nuclear Physics A} \textbf{\bibinfo{volume}{126}},
  \bibinfo{pages}{615} (\bibinfo{year}{1969}).

\bibitem[{\citenamefont{{Angulo} et~al.}(2005)\citenamefont{{Angulo},
  {Casarejos}, {Couder}, {Demaret}, {Leleux}, {Vanderbist}, {Coc}, {Kiener},
  {Tatischeff}, {Davinson} et~al.}}]{2005ApJ...630L.105A}
\bibinfo{author}{\bibfnamefont{C.}~\bibnamefont{{Angulo}}},
  \bibinfo{author}{\bibfnamefont{E.}~\bibnamefont{{Casarejos}}},
  \bibinfo{author}{\bibfnamefont{M.}~\bibnamefont{{Couder}}},
  \bibinfo{author}{\bibfnamefont{P.}~\bibnamefont{{Demaret}}},
  \bibinfo{author}{\bibfnamefont{P.}~\bibnamefont{{Leleux}}},
  \bibinfo{author}{\bibfnamefont{F.}~\bibnamefont{{Vanderbist}}},
  \bibinfo{author}{\bibfnamefont{A.}~\bibnamefont{{Coc}}},
  \bibinfo{author}{\bibfnamefont{J.}~\bibnamefont{{Kiener}}},
  \bibinfo{author}{\bibfnamefont{V.}~\bibnamefont{{Tatischeff}}},
  \bibinfo{author}{\bibfnamefont{T.}~\bibnamefont{{Davinson}}},
  \bibnamefont{et~al.}, \bibinfo{journal}{Astrophys. J. L.}
  \textbf{\bibinfo{volume}{630}}, \bibinfo{pages}{L105} (\bibinfo{year}{2005}),
  \eprint{arXiv:astro-ph/0508454}.

\bibitem[{\citenamefont{website}()}]{national}
\bibinfo{author}{\bibfnamefont{N.~A. C. R.~E.} \bibnamefont{website}},
  \urlprefix\url{http://www.nndc.bnl.gov}.

\bibitem[{\citenamefont{Voronchev and Nakao}(2003)}]{0954-3899-29-2-317}
\bibinfo{author}{\bibfnamefont{V.~T.} \bibnamefont{Voronchev}}
  \bibnamefont{and} \bibinfo{author}{\bibfnamefont{Y.}~\bibnamefont{Nakao}},
  \bibinfo{journal}{Journal of Physics G: Nuclear and Particle Physics}
  \textbf{\bibinfo{volume}{29}}, \bibinfo{pages}{431} (\bibinfo{year}{2003}),
  \urlprefix\url{http://stacks.iop.org/0954-3899/29/i=2/a=317}.

\bibitem[{\citenamefont{Fuller and Smith}(2010)}]{lepcap}
\bibinfo{author}{\bibfnamefont{G.~M.} \bibnamefont{Fuller}} \bibnamefont{and}
  \bibinfo{author}{\bibfnamefont{C.~J.} \bibnamefont{Smith}},
  \bibinfo{journal}{in preparation}  (\bibinfo{year}{2010}).

\bibitem[{\citenamefont{Schwartz et~al.}(1969)\citenamefont{Schwartz, Schrack,
  and Heaton}}]{Schwartz196936}
\bibinfo{author}{\bibfnamefont{R.~B.} \bibnamefont{Schwartz}},
  \bibinfo{author}{\bibfnamefont{R.~A.} \bibnamefont{Schrack}},
  \bibnamefont{and} \bibinfo{author}{\bibfnamefont{H.~T.}
  \bibnamefont{Heaton}}, \bibinfo{journal}{Physics Letters B}
  \textbf{\bibinfo{volume}{30}}, \bibinfo{pages}{36 } (\bibinfo{year}{1969}),
  ISSN \bibinfo{issn}{0370-2693}.

\bibitem[{\citenamefont{Clement et~al.}(1972)\citenamefont{Clement, Stoler,
  Goulding, and Fairchild}}]{Clement197251}
\bibinfo{author}{\bibfnamefont{J.~M.} \bibnamefont{Clement}},
  \bibinfo{author}{\bibfnamefont{P.}~\bibnamefont{Stoler}},
  \bibinfo{author}{\bibfnamefont{C.~A.} \bibnamefont{Goulding}},
  \bibnamefont{and} \bibinfo{author}{\bibfnamefont{R.~W.}
  \bibnamefont{Fairchild}}, \bibinfo{journal}{Nuclear Physics A}
  \textbf{\bibinfo{volume}{183}}, \bibinfo{pages}{51 } (\bibinfo{year}{1972}),
  ISSN \bibinfo{issn}{0375-9474}.

\bibitem[{\citenamefont{Larson et~al.}(1981)\citenamefont{Larson, Harvey, and
  Hill}}]{larson}
\bibinfo{author}{\bibfnamefont{D.}~\bibnamefont{Larson}},
  \bibinfo{author}{\bibfnamefont{J.~A.} \bibnamefont{Harvey}},
  \bibnamefont{and} \bibinfo{author}{\bibfnamefont{N.~W.} \bibnamefont{Hill}},
  \bibinfo{journal}{Measurement of neutron total cross sections at ORELA to 80
  MeV, ORNL-5785}  (\bibinfo{year}{1981}).

\bibitem[{Eva()}]{Evalnuc}
\urlprefix\url{http://www-nds.iaea.org/exfor/endf.htm}.

\bibitem[{\citenamefont{Kusakabe et~al.}(2009)\citenamefont{Kusakabe, Kajino,
  Boyd, Yoshida, and Mathews}}]{Kusakabe:2008kf}
\bibinfo{author}{\bibfnamefont{M.}~\bibnamefont{Kusakabe}},
  \bibinfo{author}{\bibfnamefont{T.}~\bibnamefont{Kajino}},
  \bibinfo{author}{\bibfnamefont{R.~N.} \bibnamefont{Boyd}},
  \bibinfo{author}{\bibfnamefont{T.}~\bibnamefont{Yoshida}}, \bibnamefont{and}
  \bibinfo{author}{\bibfnamefont{G.~J.} \bibnamefont{Mathews}},
  \bibinfo{journal}{Phys. Rev.} \textbf{\bibinfo{volume}{D79}},
  \bibinfo{pages}{123513} (\bibinfo{year}{2009}), \eprint{0806.4040}.

\bibitem[{\citenamefont{Kusakabe et~al.}(2008)\citenamefont{Kusakabe, Kajino,
  Boyd, Yoshida, and Mathews}}]{0004-637X-680-2-846}
\bibinfo{author}{\bibfnamefont{M.}~\bibnamefont{Kusakabe}},
  \bibinfo{author}{\bibfnamefont{T.}~\bibnamefont{Kajino}},
  \bibinfo{author}{\bibfnamefont{R.~N.} \bibnamefont{Boyd}},
  \bibinfo{author}{\bibfnamefont{T.}~\bibnamefont{Yoshida}}, \bibnamefont{and}
  \bibinfo{author}{\bibfnamefont{G.~J.} \bibnamefont{Mathews}},
  \bibinfo{journal}{The Astrophysical Journal} \textbf{\bibinfo{volume}{680}},
  \bibinfo{pages}{846} (\bibinfo{year}{2008}),
  \urlprefix\url{http://stacks.iop.org/0004-637X/680/i=2/a=846}.

\bibitem[{\citenamefont{Pospelov}(2007)}]{Pospelov:2006sc}
\bibinfo{author}{\bibfnamefont{M.}~\bibnamefont{Pospelov}},
  \bibinfo{journal}{Phys. Rev. Lett.} \textbf{\bibinfo{volume}{98}},
  \bibinfo{pages}{231301} (\bibinfo{year}{2007}), \eprint{hep-ph/0605215}.

\bibitem[{\citenamefont{Hamaguchi et~al.}(2007)\citenamefont{Hamaguchi,
  Hatsuda, Kamimura, Kino, and Yanagida}}]{Hamaguchi:2007mp}
\bibinfo{author}{\bibfnamefont{K.}~\bibnamefont{Hamaguchi}},
  \bibinfo{author}{\bibfnamefont{T.}~\bibnamefont{Hatsuda}},
  \bibinfo{author}{\bibfnamefont{M.}~\bibnamefont{Kamimura}},
  \bibinfo{author}{\bibfnamefont{Y.}~\bibnamefont{Kino}}, \bibnamefont{and}
  \bibinfo{author}{\bibfnamefont{T.~T.} \bibnamefont{Yanagida}},
  \bibinfo{journal}{Phys. Lett.} \textbf{\bibinfo{volume}{B650}},
  \bibinfo{pages}{268} (\bibinfo{year}{2007}), \eprint{hep-ph/0702274}.

\bibitem[{\citenamefont{Bird et~al.}(2008)\citenamefont{Bird, Koopmans, and
  Pospelov}}]{PhysRevD.78.083010}
\bibinfo{author}{\bibfnamefont{C.}~\bibnamefont{Bird}},
  \bibinfo{author}{\bibfnamefont{K.}~\bibnamefont{Koopmans}}, \bibnamefont{and}
  \bibinfo{author}{\bibfnamefont{M.}~\bibnamefont{Pospelov}},
  \bibinfo{journal}{Phys. Rev. D} \textbf{\bibinfo{volume}{78}},
  \bibinfo{pages}{083010} (\bibinfo{year}{2008}).

\bibitem[{\citenamefont{Kohri and Takayama}(2007)}]{PhysRevD.76.063507}
\bibinfo{author}{\bibfnamefont{K.}~\bibnamefont{Kohri}} \bibnamefont{and}
  \bibinfo{author}{\bibfnamefont{F.}~\bibnamefont{Takayama}},
  \bibinfo{journal}{Phys. Rev. D} \textbf{\bibinfo{volume}{76}},
  \bibinfo{pages}{063507} (\bibinfo{year}{2007}).

\bibitem[{\citenamefont{Kawasaki et~al.}(2007)\citenamefont{Kawasaki, Kohri,
  and Moroi}}]{Kawasaki2007436}
\bibinfo{author}{\bibfnamefont{M.}~\bibnamefont{Kawasaki}},
  \bibinfo{author}{\bibfnamefont{K.}~\bibnamefont{Kohri}}, \bibnamefont{and}
  \bibinfo{author}{\bibfnamefont{T.}~\bibnamefont{Moroi}},
  \bibinfo{journal}{Physics Letters B} \textbf{\bibinfo{volume}{649}},
  \bibinfo{pages}{436 } (\bibinfo{year}{2007}), ISSN \bibinfo{issn}{0370-2693}.

\bibitem[{\citenamefont{Kawasaki et~al.}(2008)\citenamefont{Kawasaki, Kohri,
  Moroi, and Yotsuyanagi}}]{Kawasaki:2008qe}
\bibinfo{author}{\bibfnamefont{M.}~\bibnamefont{Kawasaki}},
  \bibinfo{author}{\bibfnamefont{K.}~\bibnamefont{Kohri}},
  \bibinfo{author}{\bibfnamefont{T.}~\bibnamefont{Moroi}}, \bibnamefont{and}
  \bibinfo{author}{\bibfnamefont{A.}~\bibnamefont{Yotsuyanagi}},
  \bibinfo{journal}{Phys. Rev.} \textbf{\bibinfo{volume}{D78}},
  \bibinfo{pages}{065011} (\bibinfo{year}{2008}), \eprint{0804.3745}.

\end{thebibliography}

\end{document}